\documentclass[12pt,peerreview,onecolumn]{IEEEtranTCOM}
\pdfoutput=1
\usepackage{amsmath,epsfig,amssymb,verbatim,amsopn,subfigure,color}
\usepackage{cite,xspace}
\usepackage{array,algorithm,algorithmic}
\usepackage{bbm}
\usepackage{array}
\usepackage{multirow}
\usepackage{multicol}
\usepackage{rotating}
\usepackage{dsfont,float}
\usepackage{stfloats}
\usepackage{enumerate,makecell}
\usepackage{url}
\usepackage{tikz,graphicx}
\usepackage{booktabs}
\usepackage[font=footnotesize]{caption}
\normalsize

\ifCLASSOPTIONpeerreview
\renewcommand{\baselinestretch}{1.9}
\fi

\ifCLASSOPTIONonecolumn
    
\else
    
\fi

\newcommand{\redcom}[1]{{\color{black}#1}\xspace}

\makeatletter
\let\l@ENGLISH\l@english
\makeatother

\makeatletter
\renewcommand*{\@opargbegintheorem}[3]{\trivlist
  \item[\hskip \labelsep{\itshape #1\ #2}] {\itshape (#3):} {\normalfont}}
\makeatother
\usepackage{relsize}
\newcommand{\rmin}{r_{\textrm{min}}}
\newcommand{\rmax}{r_{\textrm{max}}}
\newcommand{\alphaL}{\alpha_\mathrm{L}}
\newcommand{\alphaN}{\alpha_\mathrm{N}}
\newcommand{\deltaL}{\delta_\mathrm{L}}
\newcommand{\deltaN}{\delta_\mathrm{N}}
\newcommand{\Gmin}{G^{\textrm{min}}}
\newcommand{\Gmax}{G^{\textrm{max}}}
\newcommand{\Pmax}{P^{\textrm{max}}}
\newcommand{\Ppt}{P_{\mathrm{PT}}}
\newcommand{\gammatr}{\gamma_{\mathrm{TR}}}
\newcommand{\gammapt}{\gamma_{\mathrm{PT}}}
\newcommand{\gL}{g_\mathrm{L}}
\newcommand{\gN}{g_\mathrm{N}}
\newcommand{\Pc}{\mathbb{P}_\mathrm{cov}}

\newtheorem{remark}{Remark}
\newtheorem{theorem}{Theorem}
\newtheorem{definition}{Definition}
\newtheorem{corollary}{Corollary}
\newtheorem{proposition}{Proposition}

\graphicspath{{figs/},{Figures/}}

\newcommand{\AuthorOne}{Xiaohui~Zhou}
\newcommand{\AuthorTwo}{Jing~Guo}
\newcommand{\AuthorThree}{Salman~Durrani}
\newcommand{\AuthorFour}{Marco~Di~Renzo}

\newcommand{\ThankOne}{X. Zhou, J. Guo and S. Durrani are with the Research School of Engineering, College of Engineering and Computer Science, The Australian National University, Canberra, ACT 2601, Australia (Emails: \{xiaohui.zhou, jing.guo, salman.durrani\}@anu.edu.au). M. Di Renzo is with the Laboratoire des Signaux et Syst\`emes, Centre National de la Recherche Scientifique, CentraleSup\'elec, University Paris Sud, Universit\'e Paris-Saclay, 91192 Gif-sur-Yvette, France (Email: marco.direnzo@lss.supelec.fr).}

\begin{document}
\title{Power Beacon-Assisted Millimeter Wave Ad Hoc Networks}
\author{{\AuthorOne,~\AuthorTwo,~\AuthorThree,~and~\AuthorFour\thanks{\ThankOne}}}
\maketitle

\begin{abstract}
Deployment of low cost power beacons (PBs) is a promising solution for dedicated wireless power transfer (WPT) in future wireless networks. In this paper, we present a tractable model for PB-assisted millimeter wave (mmWave) wireless ad hoc networks, where each transmitter (TX) harvests energy from all PBs and then uses the harvested energy to transmit information to its desired receiver. Our model accounts for realistic aspects of WPT and mmWave transmissions, such as power circuit activation threshold, allowed maximum harvested power, maximum transmit power, beamforming and blockage. Using stochastic geometry, we obtain the Laplace transform of the aggregate received power at the TX to calculate the power coverage probability. We approximate and discretize the transmit power of each TX into a finite number of discrete power levels in log scale to compute the channel and total coverage probability. We compare our analytical predictions to simulations and observe good accuracy. The proposed model allows insights into effect of system parameters, such as transmit power of PBs, PB density, main lobe beam-width and power circuit activation threshold on the overall coverage probability. The results confirm that it is feasible and safe to power TXs in a mmWave ad hoc network using PBs.
\end{abstract}

\begin{IEEEkeywords}
Wireless communications, wireless power transfer, millimeter wave transmission, power beacon, stochastic geometry.
\end{IEEEkeywords}

\newpage
\section{Introduction}
Wireless power transfer (WPT) can prolong the lifetime of low-power devices in the network and is currently in the spotlight as a key enabling technology in future wireless communication networks~\redcom{\cite{Lu-15,Tabassum-2015a,Zeng-2017}}. Compared to energy harvesting from ambient energy sources, e.g., solar, wind or ambient radio frequency (RF) sources, which may change rapidly with time, location and weather conditions, WPT has a significant advantage of being always available and controllable~\cite{Lu-15}. There are currently two main approaches to WPT: (i) simultaneous information and power transfer (SWIPT) and (ii) power beacon (PB) based approach. While SWIPT, which proposes to extract the information and power from the same signal, has been the subject of intense research in the academic community~\cite{Lu-15,Zhang-2013,ali-2013}, industry has preferred to adopt the PB approach. In this approach, low cost PBs, which do not require backhaul links, are deployed to provide dedicated power transfer in wireless networks. For example, the Cota Tile is a PB designed to wirelessly charge devices like smartphones in a home environment and was showcased at the 2017 Consumer Electronics Show (CES)~\cite{Cota-2017}.

There are two key challenges in the application of PBs to wider networks. The first challenge is the lack of tractable models for analysis and design of such networks. Although simulations can be used in this regard, exhaustive simulation of every possible scenario of interest will be extremely time-consuming and onerous. Hence, it is important to explore tractable models for PB-assisted communications in wireless networks. The second challenge is the use of practical models for WPT, which capture realistic aspects of WPT. For instance, WPT receivers (RXs) can only harvest power if the incident received power is greater than the power circuit activation threshold (typically around $-20$ dBm~\cite{Lu-15}). Similarly, WPT transmitters (TXs) have to adhere to maximum transmit power constraints due to safety considerations. Hence, it is important to adopt a realistic and practical model for WPT.


\subsection{Related Work}

\textit{\underline{Microwave (below 6 GHz) systems:}} Recently, the investigation of PBs has drawn attention in the literature from different aspects. \textit{For point-to-point or point-to-multipoint communication systems}, the resource allocation for PB-assisted system was considered in~\cite{ma-2015,zhong-2015}, where the authors mainly aimed at finding the optimum time ratio for power transfer (PT) and information transmission (IT). In~\cite{7577859}, the authors studied the PB-assisted network in the context of physical layer security, where an energy constrained source is powered by a dedicated PB. For \textit{large scale networks}, some papers have characterized the performance of PB-assisted communications using stochastic geometry, which is a powerful mathematical tool to provide tractable analysis by incorporating the randomness of users. Specifically, the feasibility of PB deployment in a cellular network, under the outage constraint at the base station, was investigated in~\cite{Huang-2014}, where cellular users are charged by PBs for uplink transmission. By considering that the secondary TX is charged by the primary user in a cognitive network, the authors derived the spatial throughput for the secondary network in~\cite{Lee-2013}. Adaptively directional PBs were proposed for sensor network in \cite{Wang-2016} and the authors found the optimal charging radius for different sensing tasks. In~\cite{Ly-2016}, three WPT schemes were proposed to select the PB for charging in a device-to-device-aided cognitive cellular network. The authors in~\cite{Guo-2015} formulated the total outage probability in a PB-assisted ad hoc network by including the energy harvesting sensitivity into the analysis. Note that all the aforementioned works considered the conventional microwave frequency band, i.e., below 6 GHz. 

\textit{\underline{MmWave systems:}} Millimeter wave (mmWave) communication, which aims to use the spectrum band typically around 30 GHz, is emerging as a key technology for the fifth generation systems~\cite{Rappaport-2013}. Considerable advancements have already been made in the understanding, modelling and analysis of mmWave communication using stochastic geometry~\cite{Andrews-2016b,Renzo-2015b,Venugopal-2015c,Maamari-2016}. From the prior work, we can summarize two distinctive features of mmWave communication: (i) owing to the smaller wavelength, mmWave allows a large number of antenna arrays with directional beamforming to be equipped at the TX and RX; (ii) since the mmWave propagation is susceptible to blockage, it causes the large difference for path-loss and fading characteristics between line of sight (LOS) and non light of sight (NLOS) environment. 

MmWave communication can be beneficial for WPT since both technologies inherently operate over short distances and the narrow beams in mmWave communication can focus the transmit power. Very recently, some papers have used stochastic geometry to analyse mmWave SWIPT networks~\cite{Khan-2016,Wang-2015}. The statistics of the aggregate received power from PBs in a mmWave ad hoc network were studied in our preliminary work in~\cite{Zhou-2017}. To the best of our knowledge, the study of a PB-assisted mmWave network using stochastic geometry, taking into account realistic and practical WPT and mmWave characteristics such as building blockages, beamforming, power circuit activation threshold, maximum harvested power and maximum transmit power, is not available in the literature.


\subsection{Our Approach and Contributions}
\redcom{In this paper, we consider a PB-assisted wireless ad hoc network under mmWave transmission where TXs adopt the harvest-then-transmit protocol, i.e., they harvest energy from the aggregate RF signal transmitted by PBs and then use the harvested energy to transmit the information to their desired RXs. Both the PT and IT phases are carried out using antenna beamforming under the mmWave channel environment, which is subjected to building blockages. Using tools from stochastic geometry, we develop a tractable analytical framework to investigate the power coverage probability, the channel coverage probability and the total coverage probability at a reference RX taking a mmWave three-state propagation model and multi-slope bounded path-loss model into account. In the proposed framework, the power coverage probability is efficiently and accurately computed by numerical inversion using the closed-form expression for the Laplace transform of the aggregate received power\footnote{\redcom{In this paper, we use the Laplace transform of a random variable to denote the Laplace transform of the distribution of a random variable for brevity.}} at the typical TX. The novel contributions of this paper are summarized as follows:
\begin{itemize}
  \item We adopt a realistic model of wirelessly powered TXs by taking into consideration (i) the power circuit activation threshold, which accounts for the minimum aggregate received power required to activate the energy harvesting circuit, (ii) the allowed maximum harvested power, which accounts for the saturation of the energy harvesting circuit and (iii) the maximum transmit power, which accounts for the safety regulation and the electrical rating of the antenna circuit.
	\item For tractable analysis of the channel coverage probability and the total coverage probability, we propose to discretize the transmit power of each TX into a finite number of discrete power levels in the log scale. Using this approximation, we derive the channel coverage probability and the total coverage probability at the typical RX. Comparison with simulation results shows that the model, with only 10 discrete levels for the transmit power of TXs, has good accuracy in the range of 5\%-10\%.
	\item Based on our proposed model, we investigate the impact of varying important system parameters (e.g., transmit power of PB, PB density, allowed maximum harvested power, directional beamforming parameters etc.) on the network performance. These trends are summarized in Table~\ref{tab:effect}.
	\item We investigate the feasibility of using PBs to power up TXs while providing an acceptable performance for IT towards RXs in mmWave ad hoc network. Our results show that under practical setups, for PB transmit power of 50 dBm and TXs with a maximum transmit power between $20-40$ dBm, which are practical and safe for human exposure, the total coverage probability is around $90\%$.
\end{itemize}}


\subsection{Notation and Paper Organization}
The following notation is used in this paper. $\Pr(\cdot)$ indicates the probability measure and $\mathbb{E}[\cdot]$ denotes the expectation operator. $j$ is the imaginary number and $\mathrm{Re}[\cdot]$ denotes the real part of a complex number. $\Gamma(x)=\int_0^\infty{t^{x-1}\exp(-t)dt}$ is the complete gamma function and $\Gamma(a,x)=\int_x^\infty{t^{a-1}\exp(-t)dt}$ is the upper incomplete gamma function, respectively. $\,_2F_1(a,b;c;z)=\frac{\Gamma(c)}{\Gamma(b)\Gamma(c-b)}\int_0^1{\frac{t^{b-1}(1-t)^{c-b-1}}{(1-tz)^a}dt}$ is the Gaussian hypergeometric function. $f_X(x)$ and $F_X(x)$ denotes the probability density function (PDF) and the cumulative distribution function (CDF) of a random variable $X$. $\mathcal{L}_X(s)=\mathbb{E}[\exp(-sX)]$ denotes the Laplace transform of a random variable $X$. A list of the main mathematical symbols employed in this paper is given in Table~\ref{tab:notation}.

\begin{table}
\centering
\caption{Summary of Main Symbols Used in the Paper.}
\label{tab:notation}
\begin{tabular}{c|c|c|c}
\hline \hline
Symbol & Definition & Symbol & Definition \\ \hline
$\phi_p$ & PB PPP & $P_p$ & PB transmit power\\ \hline
$\phi_t$ & TX PPP & $P_t$ & TX transmit power\\ \hline
$\phi_t^n$ & $n$th level TX PPP & $P_t^n$ & $n$th level TX transmit power\\ \hline
$\lambda_p$ & Density of PB PPP & $k_n$ & Portion of TXs at the $n$th level\\ \hline
$\lambda_t$ & Density of TX PPP & $N$ & Number of battery levels\\ \hline
$\lambda_t^n$ & Density of $n$th level TX PPP & $w$ & Step size of each battery level\\ \hline
$d_0$ & Length of desired TX-RX link & $\sigma^2$ & Noise power\\ \hline
$\rmin$ & Radius of the LOS region & $\eta$ & Power conversion efficiency\\ \hline
$\rmax$ & Exclusion radius of the OUT region & $\rho$ & Time switching parameter\\ \hline
$\alphaL$ & LOS link path-loss exponent & $\gammapt$ & Power circuit activation threshold\\ \hline
$\alphaN$ & NLOS link path-loss exponent & $\Pmax_1$ & Allowed maximum harvested power at active TX\\ \hline
$\gL$ & LOS link channel fading & $\Pmax_2$ & Maximum transmit power of active TX\\ \hline
$\gN$ & NLOS link channel fading & $\gammatr$ & SINR threshold\\ \hline
$m$ & Nakagami-$m$ fading parameter & $\Pc^P$ & Power coverage probability\\ \hline
$\Gmax_p$, $\Gmin_p$, $\theta_p$ & PB beamforming parameters & $\Pc^C$ & Channel coverage probability\\ \hline
$\Gmax_t$, $\Gmin_t$, $\theta_t$ & TX beamforming parameters & $\Pc$ & Total coverage probability\\ \hline
$\Gmax_r$, $\Gmin_r$, $\theta_r$ & RX beamforming parameters & {} & {}\\ \hline
\hline
\end{tabular}
\end{table}

The rest of the paper is organized as follows: Section~\ref{sec:model} describes the system model and assumptions. Section~\ref{sec:PT} focuses on the PT phase of the system and derives the power coverage probability. Section~\ref{sec:IT} details the IT phase, which covers the analysis of transmit power statistics and channel coverage probability. Section~\ref{sec:total} summaries the total coverage probability. Section~\ref{sec:result} presents the results and the effect of the system parameters on the network performance. Finally, Section~\ref{sec:conclusion} concludes the paper.

\section{System Model}\label{sec:model}
We consider a two-dimensional mmWave wireless ad hoc network, where TXs are first wirelessly charged by PBs and then they transmit information to RXs. The locations of PBs are modeled as a homogeneous Poisson point process (PPP) $\phi_p$ in $\mathbb{R}^2$ with constant node density $\lambda_p$. TXs are assumed to be randomly independently deployed and their locations are modeled as a homogeneous PPP $\phi_t$ with node density $\lambda_t$. For each TX, it has a desired RX located at a distance $d_0$ in a random direction. Throughout the paper, we use $X_i$ to denote both the random location as well as the $i$th TX itself, $Y_i$ to denote both the location and the corresponding $i$th RX and $Z_i$ to denote both the location and the $i$th PB, respectively. Note that we assume the indoor-to-outdoor penetration loss is high. Therefore, all the PBs, TXs and RXs can be regarded as outdoor devices.

\subsection{Power Transfer and Information Transmission Model}\label{sec:PTIT}
%
%
%
%

We assume that each PB has access to a dedicated power supply (e.g., a battery or power grid) and transmits with a constant power $P_p$. Time is divided into slots and let $T$ denote one time slot. Each TX adopts the harvest-then-transmit protocol to perform PT and IT. Specifically, each time slot $T$ is divided into two parts with ratio $\rho\in(0,1)$: in the first $\rho T$ seconds TX harvests energy from the RF signal transmitted by PBs \redcom{and stores the energy in an ideal (infinite capacity) battery\footnote{\redcom{In this work, we do not consider the impact of battery imperfections~\cite{Biason-2015}.}}}. In the remaining $(1-\rho)T$ seconds, TXs use all the harvested energy to transmit information to their desired RXs. \redcom{Hence, there is no interference between the PT and IT stages.} We make the following assumptions for realistic modelling of PT:
\begin{itemize}
\item Different from previous works~\cite{Huang-2014,Wang-2015,Ghazanfari-2016}, where energy harvesting activation threshold is not considered and the devices can harvest power from any amount of incident power, we assume that the TX can scavenge energy if and only if the instantaneous aggregate received power from all PBs is greater than a power circuit activation threshold $\gammapt$. If this condition is met, then the TX is called an \textit{active TX}. Otherwise, the TX will be inactive and will not scavenge any energy from the PBs.
    \item Once the energy harvesting circuit is activated, the harvested power at the active TX is assumed to be linearly proportional to the aggregate received power with power conversion efficiency $\eta$. Due to the saturation of the energy harvesting circuit, the harvested power at the active TX cannot exceed a maximum level denoted as $\Pmax_1$~\cite{Boshkovska-2015}. In addition, the active TX cannot transmit information with a power greater than $\Pmax_2$ because of the safety regulation and the electrical rating of the antenna circuit~\cite{Xia-2015}.
\end{itemize}


\subsection{MmWave Blockage Model}
Under outdoor mmWave transmissions, each link between the PB and the TX (i.e., PB-TX link) or between the TX and the RX (i.e., TX-RX link) is susceptible to building blockages due to their high diffraction and penetration characteristics~\cite{Andrews-2016b}. In this work, we adopt the state-of-the-art three-state blockage model as in~\cite{Renzo-2015b,Renzo-2016}, where each PB-TX or TX-RX link can be in one of the following three states: (i) the link is in LOS state if no blockage exists, (ii) the link is in NLOS state if blockage exists and (iii) the link is in outage (OUT) state if the link is too weak to be established.

Given that the PB-TX or TX-RX link has a length of $r$, the probabilities $p_{\mathrm{LOS}}(\cdot)$, $p_{\mathrm{NLOS}}(\cdot)$ and $p_{\mathrm{OUT}}(\cdot)$ of it being in LOS, NLOS and OUT states, respectively, are
\begin{align}\label{eq:3state}
p_{\mathrm{OUT}}(r)&=u(r-\rmax);\nonumber\\
p_{\mathrm{NLOS}}(r)&=u(r-\rmin)-u(r-\rmax);\\
p_{\mathrm{LOS}}(r)&=1-u(r-\rmin),\nonumber
\end{align}

\noindent where $u(\cdot)$ denotes the unit step function, $\rmin$ is the radius of the LOS region and $\rmax$ is the exclusion radius of the OUT region\redcom{\footnote{\redcom{Note that the two-state blockage model in \cite{Bai-2015,Renzo-2017}, which does not consider the OUT region, can be considered as a special case of the three-state blockage model with $\rmax=\infty$.}}}, as illustrated in Fig.~\ref{fig:image}. The values of $\rmin$ and $\rmax$ depend on the propagation scenario and the \redcom{mmWave} carrier frequency~\cite{Renzo-2015b}. Moreover, the communication link between TX and its desired RX is assumed to be always in LOS state.


\subsection{MmWave Channel Model}
It has been shown by the measurements that mmWave links experience different channel conditions under LOS, NLOS and OUT states~\cite{Akdeniz-2014}. Thus, we consider the following path-loss plus block fading channel model.

For the path-loss, we adopt and modify a multi-slope path-loss model~\cite{Zhang-2015} and define the path-loss of PB-TX or TX-RX link with a propagation distance of $r$ as follows
\begin{align}\label{eq:pathloss}
l(r)=\begin{cases}
					1,& 0\leqslant r<1\\
					r^{-\alphaL},& 1\leqslant r<\rmin\\
					\beta r^{-\alphaN},& \rmin\leqslant r<\rmax\\
					\infty ,& \rmax\leqslant r
				\end{cases},
\end{align}

\noindent where the first condition is added to avoid the singularity as $r\rightarrow0$, $\alphaL$ denotes the path-loss exponent for the link in LOS state, $\alphaN$ denotes the path-loss exponent for the link in NLOS state ($2\leqslant\alphaL\leqslant\alphaN$), the path-loss of the link in OUT state is assumed to be infinite \cite{Renzo-2015b} and the continuity in the multi-slope path-loss model is maintained by introducing the constant $\beta\triangleq \rmin^{\alphaN-\alphaL}$ \cite{Zhang-2015}.

As for the fading, the link under LOS state is assumed to experience Nakagami-$m$ fading, while the link under NLOS state is assumed to experience Rayleigh fading\footnote{We do not consider shadowing but it can be included using the composite fading model in~\cite{Ahmadi-2010}.}. Furthermore, both the LOS and the NLOS links experience additive white Gaussian noise (AWGN) with variance $\sigma^2$. However, under the PT phase, the AWGN power is too small to be harvested by TXs. Hence, we ignore it in the PT phase.


\subsection{Beamforming Model}
To compensate the large path-loss in mmWave band, directional beamforming is necessary for devices~\cite{Lee-2016}. In this work, we consider that mmWave antenna arrays perform directional beamforming at all PBs, TXs and RXs. Similar to~\cite{Andrews-2016b,Renzo-2015b}, the actual antenna array pattern can be approximated by a sectorized gain pattern which is given by
\begin{equation}
G_a(\theta) = \begin{cases}
\Gmax_a, & |\theta| \leq \frac{\theta_a}{2}\\
\Gmin_a, & \textrm{otherwise}
\end{cases},
\end{equation}

\noindent where subscript $a =p$ for PB, $a =t$ for TX and $a =r$ for RX, $\Gmax_a$ is the main lobe antenna gain, $\Gmin_a$ is the side lobe antenna gain, $\theta \in [-\pi, \pi)$ is the angle off the boresight direction and $\theta_a$ is the main lobe beam-width. Note that, as shown in Section~\ref{sec:gammaPT}, this model can be easily related to specific array geometries, such as an $N$ element uniform planar or linear or circular array~\cite{Venugopal-2015c}.

The main beam at the PBs are assumed to be randomly and independently oriented with respect to each other and uniformly distributed in $[-\pi, \pi)$. Given a sufficient density of the PBs, this simple strategy ensures that the aggregate received power from PBs at different locations in the network is roughly on the same order and avoids the need for channel estimation and accurate beam alignment. In addition, it has been shown in~\cite{Lee-2016} that the random directional beamforming can perform reasonably well given that more than one users need to be served.

Let $G_{ij}$ be the effective antenna gain on the link from the $i$th PB to the $j$th TX. Under sectorization, $G_{ij}$ is a discrete random variable with probability $p_k=\Pr(G_{ij}= G_k)$ and $k\in\{1,2,3,4\}$, where its distribution is summarized in Table~\ref{tab:gain_pmf}.

With regards to TX and RX, we assume that each TX points its main lobe towards its desired RX directly. Therefore, the effective antenna gain of the desired TX-RX link is $D_0=\Gmax_t\Gmax_r$ and the orientation of the beam of the interfering TX is uniformly distributed in $[-\pi, \pi)$. Let $D_{ij} (i\neq j)$ be the effective antenna gain on the link from the $i$th TX to the $j$th RX. Similar to $G_{ij}$, $D_{ij}$ is a discrete random variable with probability $q_k=\Pr(D_{ij}=D_k)$, where its distribution is given in Table~\ref{tab:gain_pmf}.

\begin{table}[t]
\begin{minipage}[b]{0.45\linewidth}
\centering
\includegraphics[width=0.65\textwidth]{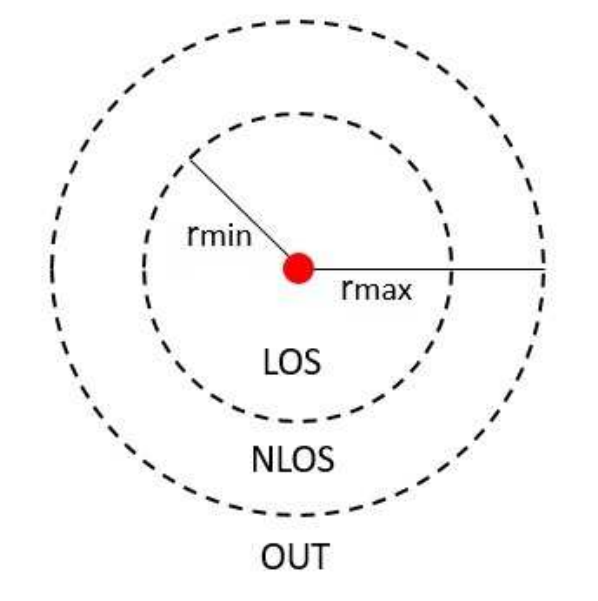}
\captionof{figure}{Illustration of mmWave blockage model.}
\label{fig:image}
\end{minipage}\hfill
\begin{minipage}[b]{0.65\linewidth}
\centering
\begin{tabular}{c|c|c|c|c}
\hline \hline
{} & \multicolumn{2}{c|}{PB-TX gain $G_{ij}$} & \multicolumn{2}{c}{TX-RX gain $D_{ij}$}\\ \hline
$k$ & Gain $G_k$ & Probability $p_k$ & Gain $D_k$ & Probability $q_k$\\ \hline \hline
1 & $\Gmax_p\Gmax_t$ & $\frac{\theta_p\theta_t}{4\pi^2}$ & $\Gmax_t\Gmax_r$ & $\frac{\theta_t\theta_r}{4\pi^2}$\\ \hline
2 & $\Gmax_p\Gmin_t$ & $\frac{\theta_p(2\pi-\theta_t)}{4\pi^2}$ & $\Gmax_t\Gmin_r$ & $\frac{\theta_t(2\pi-\theta_r)}{4\pi^2}$\\ \hline
3 & $\Gmin_p\Gmax_t$ & $\frac{(2\pi-\theta_p)\theta_t}{4\pi^2}$ & $\Gmin_t\Gmax_r$ & $\frac{(2\pi-\theta_t)\theta_r}{4\pi^2}$\\ \hline
4 & $\Gmin_p\Gmin_t$ & $\frac{(2\pi-\theta_p)(2\pi-\theta_t)}{4\pi^2}$ & $\Gmin_t\Gmin_r$ & $\frac{(2\pi-\theta_t)(2\pi-\theta_r)}{4\pi^2}$\\
\hline
\end{tabular}
\caption{Probability Mass Function of $G_{ij}$ and $D_{ij}$.}
\label{tab:gain_pmf}
\end{minipage}
\end{table}

\subsection{Metrics}\label{sec:metrics}
In this paper, we are interested in the PB-assisted mmWave wireless ad hoc network in terms of the total coverage probability for RXs (i.e., the probability that a RX can successfully receive the information from its TX after the TX successfully harvests energy from PBs). Based on the system model described above, the success of this event has to satisfy two requirements, which are:
\begin{itemize}
\item \textit{The corresponding TX is in power coverage}. Due to the random network topology and the fading channels, the aggregate received power from all PBs is a random variable. If the aggregate received power at a TX is greater than the power circuit activation threshold, the energy harvesting circuit is active and this TX can successfully harvest energy from PBs. As a result, the TX is under power coverage and IT then takes place.
\item \textit{The RX is in channel coverage}. The instantaneous transmit power for each active TX depends on its random received power. RX can receive the information from its desired TX (i.e., in channel coverage) if the signal-to-interference-plus-noise ratio (SINR) at the RX is above a certain threshold.
\end{itemize}

By leveraging the Laplace transform of the aggregate received power at a typical TX and the interference at a typical RX, we compute the power coverage probability and channel coverage probability in the following sections. In the subsequent analysis, we condition on having a reference RX $Y_0$ at the origin $(0,0)$ and its associated TX $X_0$ located at a distance $d_0$ away at $(d_0,0)$. According to Slivnyak's theorem, the conditional distribution is the same as the original one for the rest of the network~\cite{Haenggi-2013}.

\section{Power Transfer}\label{sec:PT}
In this section, we focus on the PT phase of the system. We analyze the aggregate received power at a reference TX from all PBs and find the power coverage probability at the corresponding RX.

Since the power harvested from the noise is negligible, the instantaneous aggregate received power at the typical TX $X_0$ from all the PBs can be expressed as
\begin{align}\label{eq:pagg}
\Ppt=P_p\sum_{Z_i\in\phi_p}{G_{i0} g_{i0} l(r_i)},
\end{align}

\noindent where $P_p$ is the PB transmit power, $G_{i0}$ is the effective antenna gain between $Z_i$ and $X_0$, $g_{i0}$ is the fading power gain between the $i$th PB $Z_i$ and the typical TX $X_0$, which follows the gamma distribution (under Nakagami-$m$ fading assumption) if the PB-TX link is in LOS state and exponential distribution (under Rayleigh fading distribution) if the PB-TX link is in NLOS state. $l(r_i)$ is the path-loss function given in~\eqref{eq:pathloss} and $r_i=\left|Z_i-X_0\right|$ is the Euclidean length of the PB-TX link between $Z_i$ and $X_0$. Using \eqref{eq:pagg}, the power coverage probability is defined as follows.

\begin{definition}\label{def:pcp}
The power coverage probability is the probability that the aggregate received power at the typical TX is higher than the power circuit activation threshold $\gammapt$. It can be expressed as
\begin{align}\label{eq:ppout}
\Pc^P(\gammapt)=&\Pr(\Ppt>\gammapt).
\end{align}
\end{definition}

\begin{remark}
Analytically characterizing the power coverage probability in \eqref{eq:ppout} is a challenging open problem in the literature. Generally, it is not possible to obtain a closed-form power coverage probability because of the randomness in the antenna gain, mmWave channels and locations of PBs. The closed-form expression only exists under the unbounded path-loss model with $\alpha=4$ and Rayleigh fading for all links, which is shown to be L$\acute{\textrm{e}}$vy distribution~\cite{Haenggi-2013}. To overcome this problem, some works~\cite{Bai-2015,Thornburg-2014,Khan-2016} employed the Gamma scaling method. This approach involves introducing a dummy Gamma random variable with parameter $N'$ to reformulate the original problem. However, the approach can sometimes lead to large errors with finite $N'$ value. Other works adopted the Gil-Pelaez inversion theorem~\cite{Renzo-2014} . This approach involves one fold integration and is only suitable for the random variable with a simple Laplace transform. If the Laplace transform is even moderately complicated, this method is not very efficient even if the Laplace transform is in closed-form.
\end{remark}

In this work, we adopt a numerical inversion method, which is easy to compute, \redcom{if the Laplace transform of a random variable is in closed-form,} and provides controllable error estimation. \redcom{Following~\cite{Ko-2000,Guo-2014}, the CDF of the aggregate received power $\Ppt$ is given as
\begin{subequations}
\begin{align}
F_{\Ppt}(x)&=\frac{1}{2\pi j}\int^{a+j\infty}_{a-j\infty}{\mathcal{L}_{F_{\Ppt}}(s)\exp(sx)ds}\label{eq:CDFppt1}\\
&=\frac{1}{2\pi j}\int^{a+j\infty}_{a-j\infty}{\frac{\mathcal{L}_{\Ppt}(s)}{s}\exp(sx)ds}\label{eq:CDFppt2}.
\end{align}
\end{subequations}
\noindent where \eqref{eq:CDFppt1} is obtained according to the Bromwich integral \cite{Gradshteyn-2007} and \eqref{eq:CDFppt2} follows from probability theory that $\mathcal{L}_{\Ppt}(s)=s\mathcal{L}_{F_{\Ppt}}(s)$. Using the trapezoidal rule and the Euler summation, the above integral can be transformed into a finite sum. Therefore,} we can express the power coverage probability as
\begin{align}\label{eq:ppout_mgf}
\Pc^P(\gammapt)=&1-\frac{2^{-B}\exp(\frac{A}{2})}{\gammapt}\sum_{b=0}^B\binom{B}{b}\sum_{c=0}^{C+b}\frac{(-1)^c}{D_c}\mathrm{Re}\left[\frac{\mathcal{L}_{\Ppt}(s)}{s}\right],
\end{align}
where $\mathrm{Re}[\cdot]$ is the real part operator, $s=\frac{A+j2\pi c}{2\gammapt}$, $\mathcal{L}_{\Ppt}(s)$ is the Laplace transform of $\Ppt$, $D_c=2$ (if $c=0$) and $D_c=1$ (if $c=1,2,...,C+b$). $A$, $B$ and $C$ are positive parameters used to control the estimation accuracy.

From \eqref{eq:ppout_mgf}, the \redcom{key parameter in order} to obtain the power coverage probability is $\mathcal{L}_{\Ppt}(s)$. By the definition of Laplace transform of a random variable, we express $\mathcal{L}_{\Ppt}(s)$ in closed-form in the following theorem.
\begin{theorem}\label{th:laplace_Ppt}
Following the system model in Section~\ref{sec:model}, the Laplace transform of the aggregate received power at the typical TX from all the PBs in a mmWave ad hoc network is
\begin{align}
&\mathcal{L}_{\Ppt}(s)=\!\prod_{k=1}^4\!\exp\!\Big(\!\pi\lambda_p\rmin^2p_k\!\left(m^m(m\!+\!s\rmin^{-\alphaL}P_pG_k)^{-m}\!\!-\!1\right)+\pi\lambda_p p_k\left(sP_pG_k\right)^{\deltaL}\left(\Xi_1\left(1\right)-\Xi_1\left(\rmin\right)\right)\nonumber\\
&+\pi\lambda_p p_ksP_pG_k\beta\left(\Xi_2(\rmin)-\Xi_2(\rmax)\right)+\frac{\pi\lambda_p}{2+\alphaN}p_k(sP_pG_k\beta)^{\deltaN}\left(\Xi_3(\rmin)-\Xi_3(\rmax)\right)\!\Big),\label{eq:laplaceppt}
\end{align}

where
\begin{align}
\Xi_1(r)&=\frac{m^m(r^{-\alphaL}sP_pG_k)^{-\deltaL-m}\alphaL\Gamma(1+m)}{(2+m\alphaL)\Gamma(m)}\,_2F_1\left(\!1+m,\!m+\deltaL;\!1+m+\deltaL;\!-\frac{mr^{\alphaL}}{sP_pG_k}\right),\\
\Xi_2(r)&=\frac{r^2}{r^{\alphaN}+sP_pG_k\beta},\\
\Xi_3(r)&=\frac{(r^{-\alphaN}sP_pG_k\beta)^{-\deltaN-1}}{r^{\alphaN}+sP_pG_k\beta}\Big(sP_pG_k\beta(2+\alphaN)\left.-2(r^{\alphaN}\!\!+\!\!sP_pG_k\beta)\!\,_2F_1\!\!\left(\!1,\deltaN\!+\!1;2\!+\!\deltaN;-\frac{r^{\alphaN}\beta^{-1}}{sP_pG_k}\!\right)\!\right),
\end{align}

\noindent and $\Gamma(\cdot)$ is the complete gamma function, $\,_2F_1(\cdot,\cdot;\cdot;\cdot)$ is the Gaussian (or ordinary) hypergeometric function, $\deltaL\triangleq\frac{2}{\alphaL}$ and $\deltaN\triangleq\frac{2}{\alphaN}$.
\end{theorem}
\begin{IEEEproof}
See Appendix~\ref{app:1}.
\end{IEEEproof}

By substituting \eqref{eq:laplaceppt} into \eqref{eq:ppout_mgf}, we can compute the power coverage probability. As shown in Theorem~\ref{th:laplace_Ppt}, the Laplace transform of $\Ppt$ is in closed-form; hence, $\Pc^P(\gammapt)$ is just a summation over a finite number of terms. Following the selection guideline of parameters $A$, $B$ and $C$ in \cite{Guo-2014}, we can achieve a stable numerical result by carefully choosing them.

\begin{figure}[t]
\centering
\includegraphics[width=0.5\textwidth]{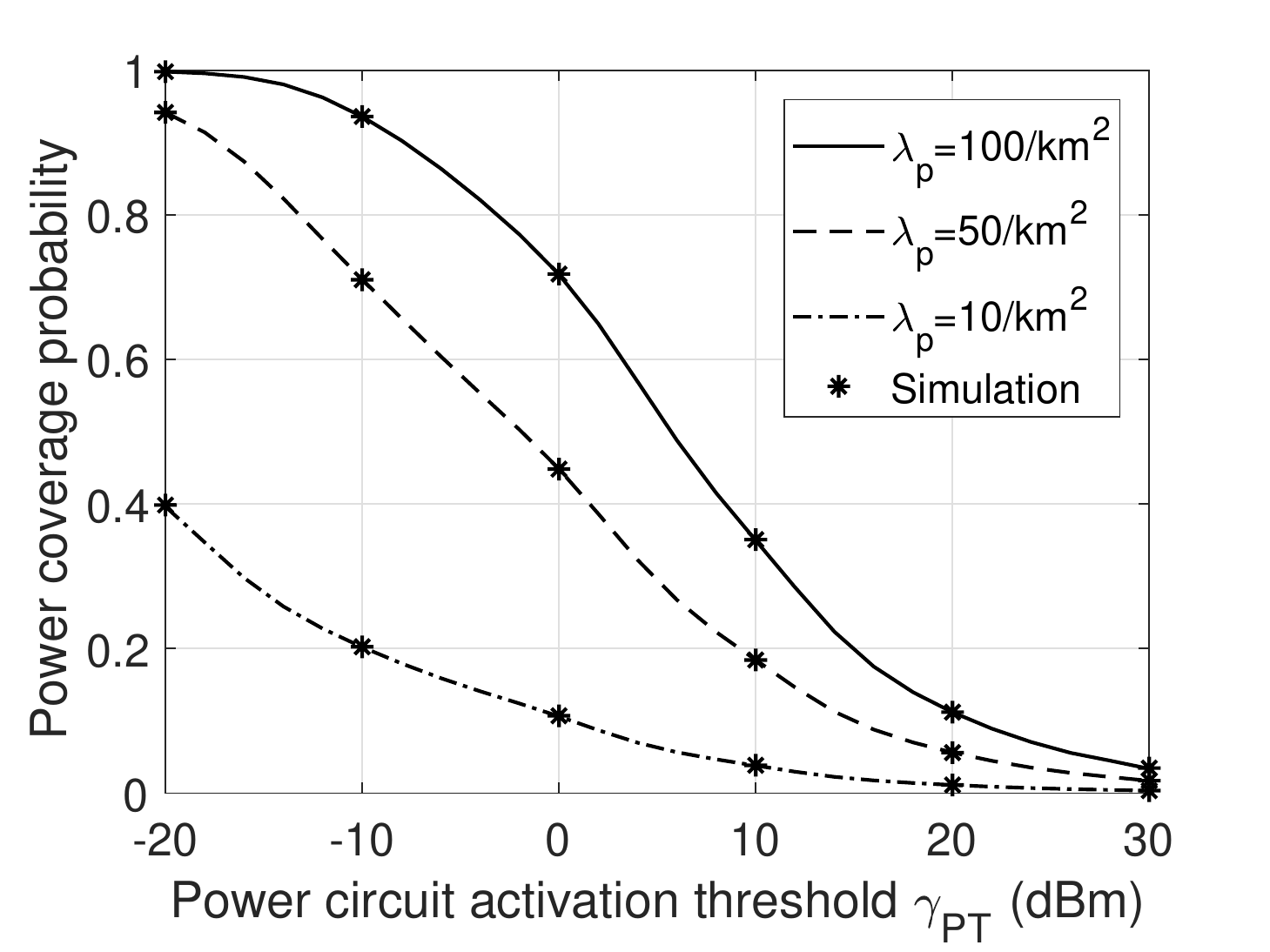}
\centering
\vspace{-3mm}
\caption{Power coverage probability versus power circuit activation threshold $\gammapt$ \redcom{for different PB densities}. Other system parameters follow Table~\ref{tab:values}.}
\centering
\label{fig:ppout}
\end{figure}

Before ending this section, we validate the analysis for the power coverage probability. Fig.~\ref{fig:ppout} plots the power coverage probability versus power circuit activation threshold. The simulation results are generated by averaging over $10^8$ Monte Carlo simulation runs. We set $A=24$, $B=20$ and $C=30$ in order to achieve an estimation error of $10^{-10}$. The other system parameters follow Table~\ref{tab:values}. From the figure, we can see that the analytical results match perfectly with the simulation results, which demonstrates the accuracy of the proposed approach. \redcom{Fig.~\ref{fig:ppout} also shows that the power coverage probability increases with the density of PBs, because the aggregate received power at TX increases as the PB density increases.} 
%
%

\section{Information Transmission}\label{sec:IT}
In this section we focus on the IT phase between the TX and RX. We assume that the TX uses all the harvested energy in the IT phase. As indicated in Section~\ref{sec:PTIT}, the transmit power of an active TX is a random variable which depends on its harvested power. Hence, we first evaluate the transmit power for an active TX. Then, we calculate the channel coverage probability at the reference RX. Note that the derived channel coverage probability is in fact a conditional probability, which is conditioned on the reference TX-RX link being active.

\subsection{Transmit Power and Locations of Active TX}\label{sec:ptx}

Using the PT assumptions in Section~\ref{sec:PTIT}, the instantaneous transmit power for each active TX is
\begin{align}\label{eq:powertx}
P_t=\begin{cases}
				\eta\frac{\rho}{1-\rho}\Ppt,& \min(\eta^{-1}\Pmax_1,\frac{1-\rho}{\eta\rho}\Pmax_2)>\Ppt\geqslant\gammapt\\
				\min(\frac{\rho}{1-\rho}\Pmax_1,\Pmax_2), & \Ppt\geqslant \min(\eta^{-1}\Pmax_1,\frac{1-\rho}{\eta\rho}\Pmax_2)
		\end{cases},	
\end{align}

\noindent where $0\leqslant\eta\leqslant1$ is the power conversion efficiency. Note that the first condition in \eqref{eq:powertx} comes from the fact that the received power at an active TX must be greater than $\gamma_{\mathrm{PT}}$. For the second condition in \eqref{eq:powertx}, $\Pmax_1$ is the maximum harvested power at an active TX when the energy harvesting circuit is saturated and $\Pmax_2$ is the maximum transmit power for an active TX. Thus, the second condition caps the transmit power by the allowed maximum harvested power constraint or the maximum transmit power constraint.

The following remarks discuss the modelling challenges and proposed solution for characterizing $P_t$.
\begin{remark}
To the best of our knowledge, the closed-form expression for the PDF of $P_t$ is very difficult to obtain. This is because $P_t$ and $\Ppt$ are correlated and the closed-form CDF of $\Ppt$ is not available according to Section~\ref{sec:PT}. In the literature, some papers~\cite{Wang-2015,Akbar-2016} have proposed to use the average harvested power as the transmit power for each TX. However, this does not always lead to accurate results. Hence, inspired from the approach in~\cite{Sakr-2015b}, we propose to discretize $P_t$ in \eqref{eq:powertx} into a finite number of levels. We show that this approximation allows tractable computation of the channel coverage probability. The accuracy of this approximation depends on the number of levels. Our results in Section~\ref{sec:validate} show that if we discretize the power level in the log scale, a reasonable level of accuracy is reached with as little as $10$ levels.
\end{remark}

\begin{remark}\label{rm:discrete}
From \eqref{eq:powertx}, we can see that $P_t$ depends on $\Ppt$. Hence, the motivation for discretizing $P_t$ in the log scale comes from looking into two important measures of $\Ppt$, the skewness and the kurtosis. The skewness and the kurtosis describe the shape of the probability distribution of $\Ppt$. As presented in \cite{Zhou-2017}, the distribution of the aggregate received power is skewed to the right with a heavy tail, because both the skewness and the kurtosis of $\Ppt$ are much greater than 0 for most cases. Therefore, most of the TXs will be at the lowest power level if we discretize $P_t$ in linear scale. Hence, we discretize the power level in the log scale. This improves the accuracy of the approximation.
\end{remark}

Let $N+1$ and $w$ denote the total number of levels and the step size of each level, respectively. They are related by $w=\left(\frac{\min\left(\eta^{-1}\Pmax_1,\frac{1-\rho}{\eta\rho}\Pmax_2\right)-\gammapt}{N}\right)$dBm. We further define $k_n$ as the portion of TXs whose $P_t$ is at the $n$th level, i.e., $k_n=\Pr\left(\left(nw+\gammapt\right)\mathrm{dBm}\leqslant \Ppt<\left(\left(n+1\right)w+\gammapt\right)\mathrm{dBm}\right)$ for $n=\{0,1,2,...,N-1\}$ and $k_N=\Pr(\Ppt\geqslant \min(\eta^{-1}\Pmax_1,\frac{1-\rho}{\eta\rho}\Pmax_2))$. Combining with the power coverage probability derived in Section~\ref{sec:PT}, we can express $k_n$ as
\begin{align}\label{eq:kn}
k_n=\begin{cases}
				\Pc^P\left((nw+\gammapt)\mathrm{dBm}\right)-\Pc^P\left(\left(\left(n+1\right)w+\gammapt\right)\mathrm{dBm}\right), & n=\{0,1,2,...,N-1\}\\
				\Pc^P(\min(\eta^{-1}\Pmax_1,\frac{1-\rho}{\eta\rho}\Pmax_2)), & n=N
		\end{cases}.
\end{align}

The above expression allows us to determine the portion of TXs whose $P_t$ is at the $n$th level. The transmit power for the active TX at the $n$th level is
\begin{align}\label{eq:Ptx}
P^n_t=\left(\eta\frac{\rho}{1-\rho}10^\frac{nw+\gamma_{\mathrm{PT}}-30}{10}\right)\mathrm{W}.
\end{align}

The next step is to decide how to model the locations of the TXs whose $P_t$ is at the $n$th level. \redcom{This is discussed in the remark below.
\begin{remark}\label{rm:correlation}
In general, the location and the transmit power of an active TX are correlated, i.e., a TX has higher chance to be activated and transmits with a larger power, if its location is closer to a PB. However, it is not easy to identify and fit a spatial point process with local clustering to model the location of active TXs \cite{Guo-2015, Sakr-2015}. In this paper, for analytical tractability, we assume that the location and the transmit power of an active TX are independent, i.e., a TX in $\phi_t$ can have a transmit power of $P^n_t$ with probability $k_n$ independently of other TXs.
\end{remark}}

Therefore, using the thinning theorem, we interpret the active TX at the $n$th level as an independent homogeneous PPP with node density $\lambda^n_t=k_n\lambda_t$, denoted as $\phi^n_t$. The accuracy of this approximation will be validated in Section~\ref{sec:validate}.

\subsection{Channel Coverage Probability}
Given that the desired TX is active, the instantaneous SINR at the reference RX, $Y_0$, is given as
\begin{align}\label{eq:sinr}
SINR=\frac{P_{X_0} D_0 h_0 l(d_0)}{\sum_{X_i\in\phi_\textrm{active}}P_{X_i} D_{i0} h_{i0} l(X_i)+\sigma^2},
\end{align}

\noindent where $h_0$ and $h_{i0}$ denote the fading power gains on the reference link and the $i$th interference link respectively, $D_0$ and $D_{i0}$ denote the beamforming antenna gain at the RX from its reference TX and the $i$th interfering TX respectively and $\sigma^2$ is the AWGN power. $P_{X_0}$ and $P_{X_i}$ are the transmit power for the reference TX and the active TX $X_i$, respectively. Using \eqref{eq:sinr}, the power coverage probability is defined as follows.

\begin{definition}
The channel coverage probability is the probability that the SINR at the reference RX is above a threshold $\gammatr$ and can be expressed as
\begin{align}\label{eq:pcout}
\Pc^C(\gammatr)=&\Pr(SINR>\gammatr).
\end{align}
\end{definition}

\begin{remark}
It is possible to employ the numerical inversion method in Section~\ref{sec:PT} to find the channel coverage probability. In doing so, the Laplace transform of the term $\frac{I_X+\sigma^2}{P_{X_0} D_0 h_0 l(d_0)}$ is required. This Laplace tranform cannot be expressed in closed-form because of the random variables $P_{X_0}$ and $h_0$ in the denominator. Although it is still computable, it leads to greater computation complexity. Consequently, we employ the reference link power gain (RLPG) based method in \cite{Guo-2014} to efficiently find the channel coverage probability. The basic principle of this approach is to first find the conditional outage probability in terms of the CDF of the reference link’s fading power gain and then remove the conditioning on the fading power gains and locations of the interferers, respectively. In order to apply this method, the reference TX-RX link is assumed to undergo Nakagami-$m$ fading with integer $m$. The result for the conditional channel coverage probability is presented in the following proposition.
\end{remark}

\begin{proposition}\label{pr:pcout}
Following the system model in Section~\ref{sec:model}, the conditional channel coverage probability at the reference RX in a mmWave ad hoc network is
\begin{align}\label{eq:pcout_rlpg}
\Pc^C(\gammatr)=&\sum_{n=0}^N\sum_{l=0}^{m-1}\frac{(-s)^l}{l!}\frac{dl}{ds^l}\mathcal{L}_{I_X+\sigma^2}(s)\frac{k_n}{\Pc^P(\gammapt)},
\end{align}
where $I_X=\sum_{n=0}^N\sum_{X_i\in\phi^n_t}P^n_t D_{i0} h_{i0} l(X_i)$ and $s=\frac{m\gammatr}{P_t^nD_0l(d_0)}$.
\end{proposition}
\begin{IEEEproof}
See Appendix~\ref{app:2}.
\end{IEEEproof}

\eqref{eq:pcout_rlpg} needs the Laplace transform of the interference plus noise. Using stochastic geometry, we can derive it and the result is shown in the following corollary.
\begin{corollary}\label{co:laplace_I}
Following the system model in Section~\ref{sec:model} and the discretization assumption in Section~\ref{sec:ptx}, the Laplace transform of the aggregate interference plus noise at the reference RX in a mmWave ad hoc network is
\begin{align}
\mathcal{L}_{I_X+\sigma^2}(s)=&\prod_{n=0}^N\!\prod_{k=1}^4\!\exp\!\Big(\!\pi\lambda_t^n\rmin^2q_k\!\left(m^m(m\!+\!s\rmin^{-\alphaL}P_t^nD_k)^{-m}\!\!-\!1\right)+\pi\lambda_t^n q_k\left(sP_t^nD_k\right)^{\deltaL}\left(\Xi'_1\left(1\right)-\Xi'_1\left(\rmin\right)\right)\nonumber\\
+&\pi\lambda_t^nq_ksP_t^nD_k\beta\left(\Xi'_2(\rmin)-\Xi'_2(\rmax)\right)+\frac{\pi\lambda_t^n}{2+\alphaN}q_k(sP_t^nD_k\beta)^{\deltaN}\left(\Xi'_3(\rmin)-\Xi'_3(\rmax)\right)\!\Big)\exp(-s\sigma^2),\label{eq:laplace_IN}
\end{align}
\noindent where
\begin{align}
\Xi'_1(r)&=\frac{m^m(r^{-\alphaL}sP_t^nD_k)^{-\deltaL-m}\alphaL\Gamma(1+m)}{(2+m\alphaL)\Gamma(m)}\,_2F_1\left(\!1+m,\!m+\deltaL;\!1+m+\deltaL;\!-\frac{mr^{\alphaL}}{sP_t^nD_k}\right),\\
\Xi'_2(r)&=\frac{r^2}{r^{\alphaN}+sP_t^nD_k\beta},\\
\Xi'_3(r)&=\frac{(r^{-\alphaN}sP_t^nD_k\beta)^{-\deltaN-1}}{r^{\alphaN}+sP_t^nD_k\beta}\Big(sP_t^nD_k\beta(2+\alphaN)\left.-2(r^{\alphaN}\!\!+\!\!sP_t^nD_k\beta)\!\,_2F_1\!\!\left(\!1,\deltaN\!+\!1;2\!+\!\deltaN;-\frac{r^{\alphaN}\beta^{-1}}{sP_t^nD_k}\!\right)\!\right).
\end{align}
\end{corollary}
\begin{IEEEproof}
Following the definition of Laplace transform, we have
\begin{align}
\mathcal{L}_{I_X+\sigma^2}(s)=&\mathbb{E}_{I_X}[\exp(-s(I_X+\sigma^2))]=\mathbb{E}_{I_X}[\exp(-sI_X)]\exp(-s\sigma^2)=\mathcal{L}_{I_X}(s)\exp(-s\sigma^2),
\end{align}
where the Laplace transform of the aggregate interference can be expressed as
\ifCLASSOPTIONpeerreview
\begin{align}
\mathcal{L}_{I_X}(s)=&\mathbb{E}_{I_X}[\exp(-sI_X)]=\mathbb{E}_{D_{i0},h_{i0},\phi^n_t}\left[\exp\left(-s\sum_{n=0}^N\sum_{X_i\in\phi^n_t}P^n_t D_{i0} h_{i0} l(X_i)\right)\right]\nonumber\\
=&\prod_{n=0}^N\mathbb{E}_{D_{i0},h_{i0},\phi^n_t}\left[\exp\left(-s\sum_{X_i\in\phi^n_t}P^n_t D_{i0} h_{i0} l(X_i)\right)\right].\label{eq:laplace_I}
\end{align}
\else
\begin{align}
\mathcal{L}_{I_X}(s)=&\mathbb{E}_{I_X}[\exp(-sI_X)]\nonumber\\
=&\mathbb{E}_{D_{i0},h_{i0},\phi^n_t}\left[\exp\left(-s\sum_{n=0}^N\sum_{X_i\in\phi^n_t}P^n_t D_{i0} h_{i0} l(X_i)\right)\right]\nonumber\\
=&\prod_{n=0}^N\mathbb{E}_{D_{i0},h_{i0},\phi^n_t}\left[\exp\left(-s\sum_{X_i\in\phi^n_t}P^n_t D_{i0} h_{i0} l(X_i)\right)\right].\label{eq:laplace_I}
\end{align}
\fi

\noindent Then, following the same steps as the proof of Laplace transform of aggregate received power in Appendix~\ref{app:1}, we can find the expectation in~\eqref{eq:laplace_I} and arrive at the result in \eqref{eq:laplace_IN}.
\end{IEEEproof}

The Laplace transform shown in Corollary~\ref{co:laplace_I} is in closed-form. Substituting \eqref{eq:laplace_IN} into \eqref{eq:pcout_rlpg}, we can easily compute the conditional channel coverage probability. Note that \eqref{eq:pcout_rlpg} requires higher order derivatives of the Laplace transform of the interference plus noise $\frac{dl}{ds^l}\mathcal{L}_{I_X+\sigma^2}(s)$, which can be yielded in closed-form using chain rules and changing variables. For brevity, the details are omitted here. 

\section{Total Coverage Probability}\label{sec:total}
As discussed in Section~\ref{sec:metrics}, the event that the information can be successfully delivered to RX has two requirements, i.e., satisfying power coverage and channel coverage. Based on our definition, the total coverage probability is
\begin{align}
\Pc(\gammapt,\gammatr)=&\Pr(\textrm{TX is in power coverage \& RX is in channel coverage})\nonumber\\
=&\Pr(\textrm{TX is in power coverage})\Pr(\textrm{RX is in channel coverage}\mid \textrm{TX is in power coverage})\nonumber
\end{align}
Combining our analysis presented in Section~\ref{sec:PT} and \ref{sec:IT}, we have
\begin{align}\label{eq:pout}
\Pc(\gamma_{\mathrm{PT}},\gamma_{\mathrm{TR}})=&\Pc^P(\gamma_{\mathrm{PT}})\Pc^C(\gamma_{\mathrm{TR}})\nonumber\\
=&\sum_{n=0}^N\sum_{l=0}^{m-1}\frac{(-s)^l}{l!}\frac{dl}{ds^l}\mathcal{L}_{I_X+\sigma^2}(s)k_n,
\end{align}
where $s=\frac{m\gammatr}{P_t^nD_0l(d_0)}$, $\mathcal{L}_{I_X+\sigma^2}(s)$ is given in Corollary~\ref{co:laplace_I}, $k_n$ is presented in~\eqref{eq:kn}, which is determined by the power coverage probability. The key metrics are summarized in Table~\ref{tab:metrics}.

\begin{table}[t]
\centering
\caption{Summary of the Analytical Model for PB-assisted mmWave Ad Hoc Networks.}
\label{tab:metrics}
\begin{tabular}{c|c|c}
\hline
Performance metrics & General form & Key factor(s)\\ \hline
Power coverage probability & \eqref{eq:ppout_mgf} & $\mathcal{L}_{\Ppt}(s)$ in~\eqref{eq:laplaceppt}\\ \hline
Channel coverage probability & \eqref{eq:pcout_rlpg} & $\Pc^P(\gammapt)$ in~\eqref{eq:ppout_mgf} \& $\mathcal{L}_{I_X+\sigma^2}(s)$ in~\eqref{eq:laplace_IN} \\ \hline
Total coverage probability & \eqref{eq:pout} & $\Pc^P(\gammapt)$ in~\eqref{eq:ppout_mgf} \& $\mathcal{L}_{I_X+\sigma^2}(s)$ in~\eqref{eq:laplace_IN}\\
\hline
\end{tabular}
\end{table}

\section{Results}\label{sec:result}
In this section, we first validate the proposed model and then discuss the design insights provided by the model. Unless stated otherwise, the values of the parameters summarized in Table~\ref{tab:values} are used. The chosen values are consistent with the literature in mmWave and WPT~\cite{Andrews-2016b,Lu-15,Renzo-2015b}. \redcom{Note that the values of $\rmin$ and $\rmax$ correspond to 28 GHz mmWave carrier frequency~\cite{Andrews-2016b}.} We mainly focus on illustrating the results for total coverage probability and channel coverage probability. As for the power coverage probability, it will be explained within the text.


Table~\ref{tab:effect} summarizes the impact of varying the important system parameters\redcom{\footnote{\redcom{Note that the trends reported in Table~\ref{tab:effect} remain the same for a two-state blockage model.}}}, i.e., SINR threshold $\gammatr$, PB density $\lambda_p$, TX density $\lambda_t$, PB transmit power $P_p$, radius of the LOS region $\rmin$, power circuit activation threshold $\gammapt$, the beam-width of the main lobe of TX $\theta_t$, RX's main lobe beam-width $\theta_r$, allowed maximum harvested power at active TX $\Pmax_1$, time switching parameter $\rho$ and TX maximum transmit power $\Pmax_2$ on the three network performance metrics. In Table~\ref{tab:effect}, $\uparrow$, $\downarrow$ and - denote increase, decrease and unrelated, respectively. $\uparrow\downarrow$ represents that the performance metric first increases then decreases with the system parameter. Please note that the trends in Table~\ref{tab:effect} originate from the analysis of the numerical results, which is presented in detail in the following subsections.
\ifCLASSOPTIONpeerreview
\begin{table}
\centering
\caption{Parameter Values.}
\label{tab:values}
\begin{tabular}{|c|c|c|c|c|c|c|c|}
\hline
Parameter & Value & Parameter & Value & Parameter & Value & Parameter & Value\\ \hline
$\lambda_p$ & 50 /km$^2$ & $m$ & 5 & $\alphaL$ & 2 & $\rho$ & 0.5\\ \hline
$\lambda_t$ & 100 /km$^2$ & $\Gmax_p$, $\Gmin_p$, $\theta_p$ & [20 dB, $-$10 dB, $30^\textrm{o}$] & $\alphaN$ & 4 & $\eta$ & 0.5\\ \hline
$d_0$ & 20 m & $\Gmax_t$, $\Gmin_t$, $\theta_t$ & [10 dB, $-$10 dB, $45^\textrm{o}$] & $P_p$ & 40 dBm & $\gammapt$ & -20 dBm\\ \hline
$\rmin$ & 100 m & $\Gmax_r$, $\Gmin_r$, $\theta_r$ & [10 dB, $-$10 dB, $45^\textrm{o}$] & $\Pmax_1$ & 20 dBm & $\gammatr$ & 30 dBm\\ \hline
$\rmax$ & 200 m & $\sigma^2$ & -30 dBm & $\Pmax_2$ & 30 dBm & $N$ & 10\\ \hline
\end{tabular}
\end{table}
\else
\begin{table}
\centering
\caption{Parameter Values.}
\label{tab:values}
\begin{tabular}{|c|c|c|c|}
\hline
Parameter & Value & Parameter & Value\\ \hline
$\lambda_p$ & 50 /km$^2$ & $m$ & 5\\ \hline
$\lambda_t$ & 100 /km$^2$ & $\Gmax_p$, $\Gmin_p$, $\theta_p$ & [20 dB, $-$10 dB, $30^\textrm{o}$] \\ \hline
$d_0$ & 20 m & $\Gmax_t$, $\Gmin_t$, $\theta_t$ & [10 dB, $-$10 dB, $45^\textrm{o}$]\\ \hline
$\rmin$ & 100 m & $\Gmax_r$, $\Gmin_r$, $\theta_r$ & [10 dB, $-$10 dB, $45^\textrm{o}$]\\ \hline
$\rmax$ & 200 m & $\sigma^2$ & -30 dBm \\ \hline
$\alphaL$ & 2 & $\rho$ & 0.5 \\ \hline
$\alphaN$ & 4 & $\eta$ & 0.5 \\ \hline
$P_p$ & 40 dBm & $\gammapt$ & -20 dBm \\ \hline
$\Pmax_1$ & 20 dBm & $\gammatr$ & 30 dBm \\ \hline
$\Pmax_2$ & 30 dBm & $N$ & 10 \\ \hline
\end{tabular}
\end{table}
\fi
\begin{table}
\centering
\caption{Effect of Important System Parameters.}
\label{tab:effect}
\begin{tabular}{|c|c|c|c|}
\hline
Parameter & Power coverage probability  & Channel coverage probability & Total coverage probability\\ \hline
Increasing $\gammatr$ & - & $\downarrow$ & $\downarrow$\\ \hline
Increasing $\lambda_p$ & $\uparrow$ & $\downarrow\uparrow$ & $\uparrow$ \\ \hline
Increasing $\lambda_t$ & - & $\downarrow$ & $\downarrow$ \\ \hline
Increasing $P_p$ & $\uparrow$ & $\uparrow$ & $\uparrow$ \\ \hline
Increasing $\rmin$ & $\uparrow$ & $\uparrow$ & $\uparrow$ \\ \hline
Increasing $\gammapt$ & $\downarrow$ & $\uparrow$ & $\downarrow$\\ \hline
Increasing $\theta_t$ and $\theta_r$  & $\uparrow$ & $\downarrow$ & $\downarrow$ \\ \hline
Increasing $\Pmax_1$ & - & $\uparrow\downarrow$ & $\uparrow\downarrow$\\ \hline
Increasing $\rho$ & - & $\uparrow$ & $\uparrow$\\ \hline
Increasing $\Pmax_2$ & - & $\downarrow$ & $\downarrow$ \\ \hline
\end{tabular}
\end{table}

\begin{figure*}[t]
\centering
\subfigure[$\lambda_p=50$ /km$^2$, $\lambda_t=500$ /km$^2$.]{\includegraphics[scale=0.5]{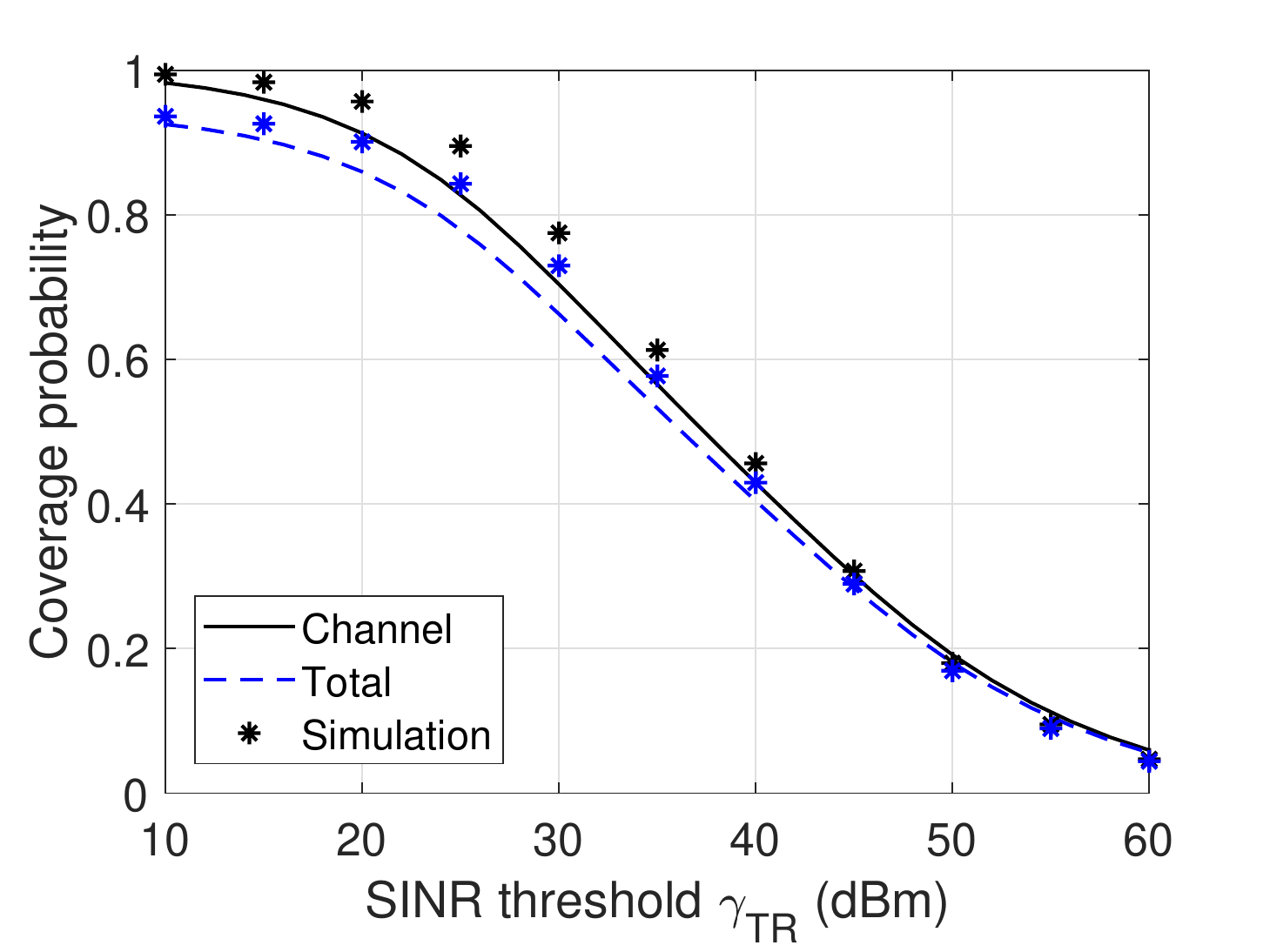}}
\subfigure[$\lambda_p=10$ /km$^2$, $\lambda_t=100$ /km$^2$.]{\includegraphics[scale=0.5]{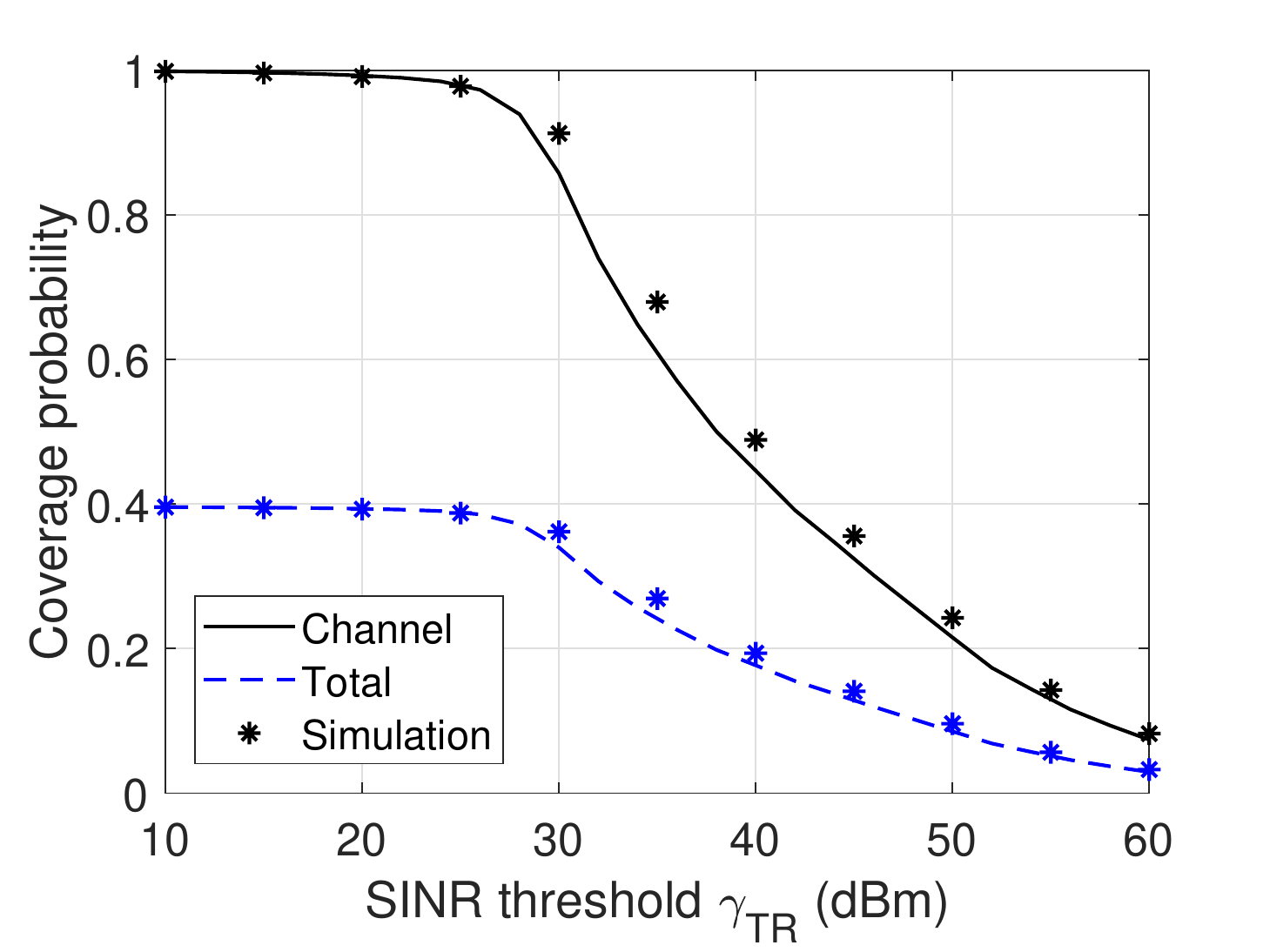}}
\subfigure[$\lambda_p=50$ /km$^2$, $\lambda_t=250$ /km$^2$.]{\includegraphics[scale=0.5]{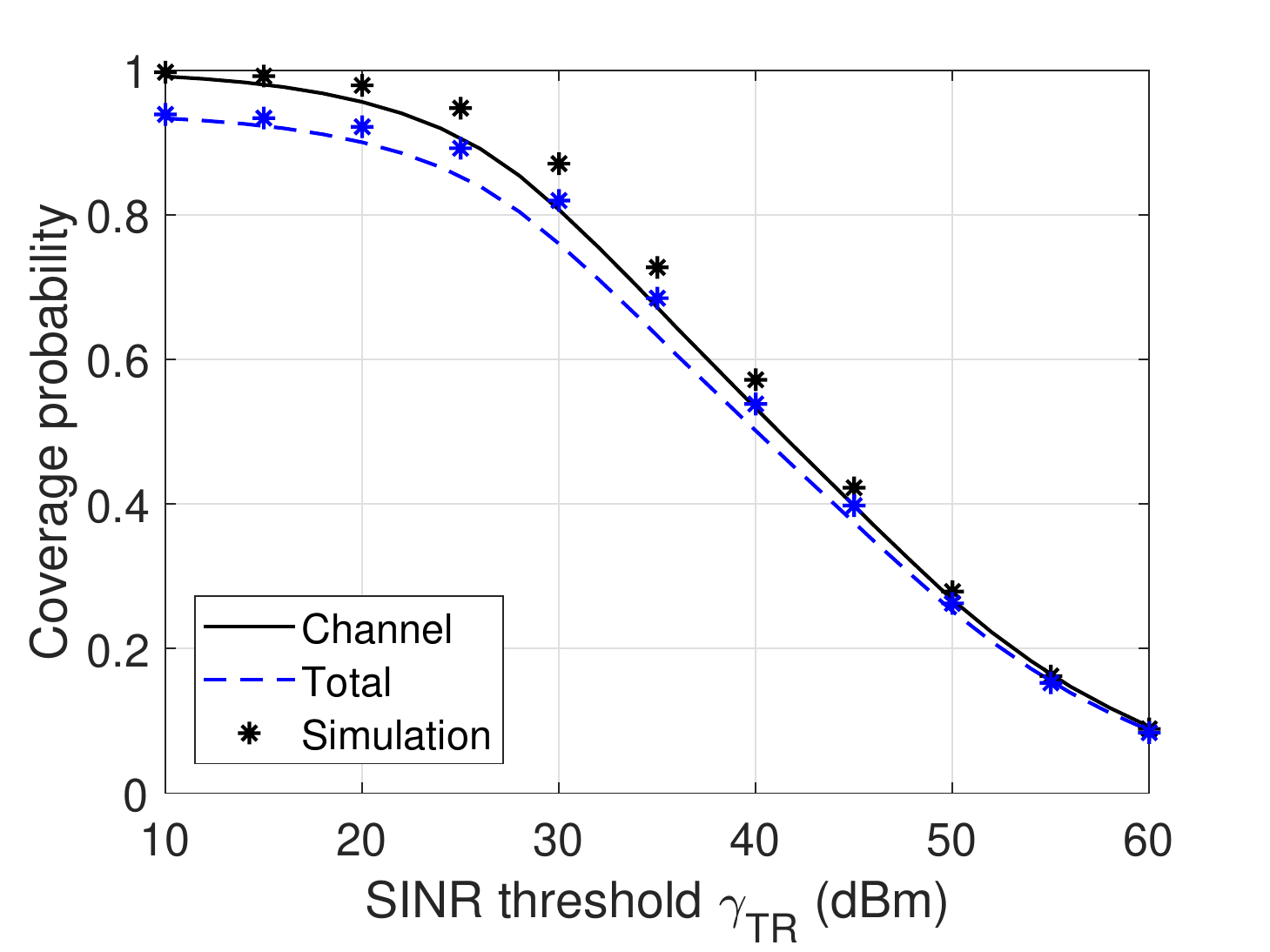}}
\subfigure[$\lambda_p=10$ /km$^2$, $\lambda_t=50$ /km$^2$.]{\includegraphics[scale=0.5]{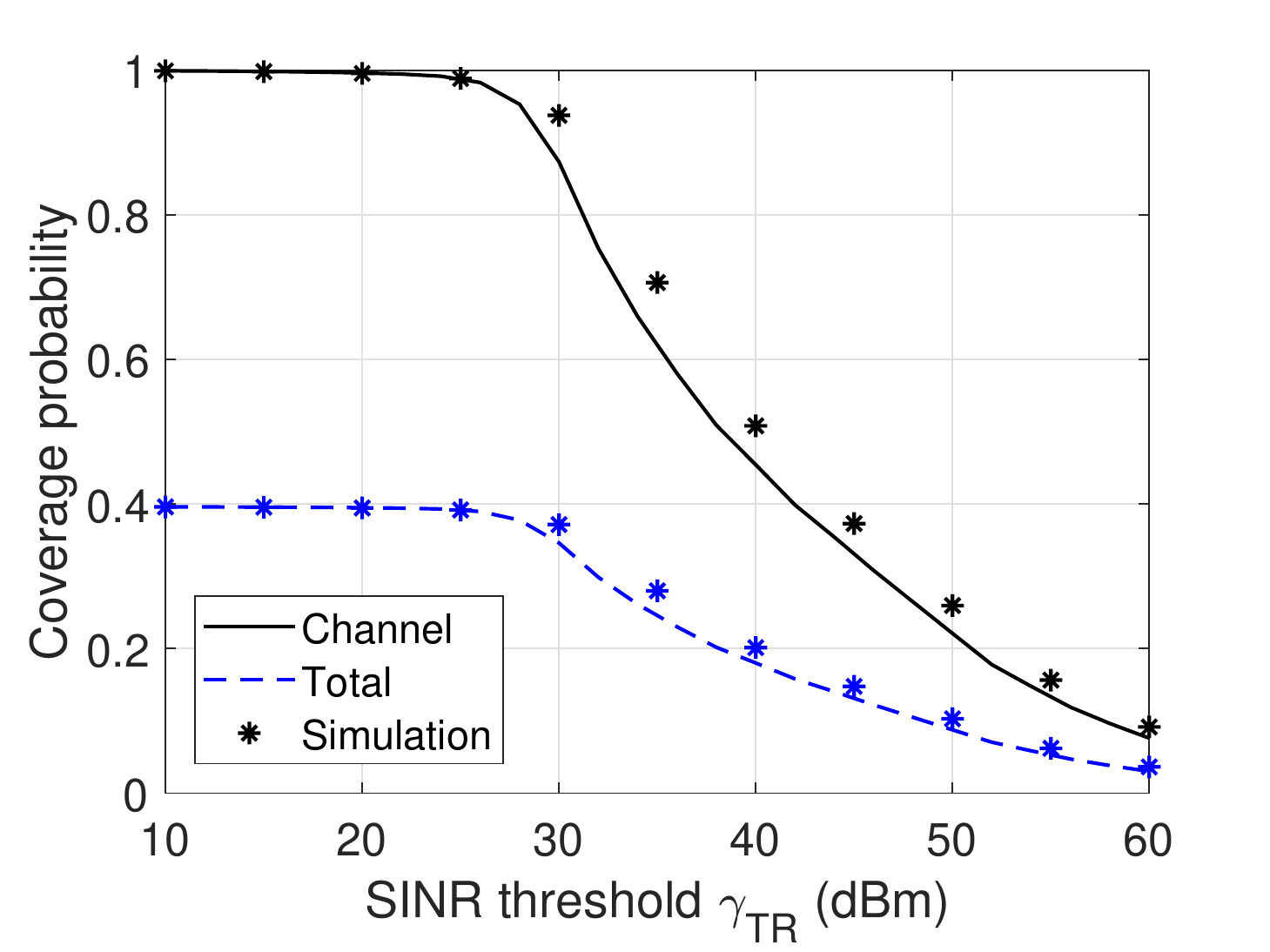}}
\vspace{-3mm}
\caption{Channel coverage probability and total coverage probability versus SINR threshold $\gammatr$. The PB density is 50 and 10 per $\textrm{km}^2$ and the TX density is 500, 100, 250 and 50 per $\textrm{km}^2$.}
\label{fig:pcout}
\end{figure*}

\subsection{Model Validation}\label{sec:validate}
In this section, we validate the proposed model for the channel coverage probability and the total coverage probability. Fig.~\ref{fig:pcout} plots the channel coverage probability and the total coverage probability for a reference RX against SINR threshold for different densities of PBs and TXs. The analytical results are obtained using Proposition~\ref{pr:pcout} and \eqref{eq:pout} with 10 discrete levels for $P_t$. The simulation results are generated by averaging over $10^8$ Monte Carlo simulation runs and do not assume any discretization of power levels.

From the figure, we can see that our analytical results provide a good approximation to the simulation. The small gap between them comes from two reasons: (i) discretization of the power levels, \redcom{as discussed in Remark~\ref{rm:discrete}}, and (ii) ignorance of the correlation between the location and the transmit power of active TX, \redcom{as discussed in Remark~\ref{rm:correlation}}. From Fig.~\ref{fig:pcout}, we can see that the gap between the simulation and the analytical results is smaller, when $\gammatr$ is higher. At $\gammatr=30$ dBm, which is a typical SINR threshold, the relative errors between the proposed model and the simulation results for both channel coverage probability and total coverage probability are between $5\%$ to $10\%$. This validates the use of 10 discrete levels for $P_t$, which provides good accuracy.

\textit{Insights:} Comparing the four cases for the different PB and TX densities, Fig.~\ref{fig:pcout} shows that: (i) The channel coverage probability decreases while the total coverage probability increases as PB density increases. As the PB density increases, the aggregate received power at TX increases as well as the number of active TXs. Therefore, interfering power received by the RX is higher and the channel coverage probability decreases. However, the total coverage probability increases because the power coverage probability increases with the PB density. (ii) When the PB density is low, the TXs are very likely to be inactive and the total coverage probability is dominated by the power coverage probability. When the PB density is high, the TXs are very likely to be active. Hence, the interference is strong and the channel coverage probability dominates the total coverage probability. (iii) For the same PB density, both the total coverage probability and the channel coverage probability are higher, when the TX density is lower. This is because more interfering TXs exist if TX density increases.


\subsection{Effect of PB Transmit Power}\label{sec:Pp}
\begin{figure*}[t]
\centering
\subfigure[Channel coverage and total coverage with different $\rmin$.]{\includegraphics[scale=0.6]{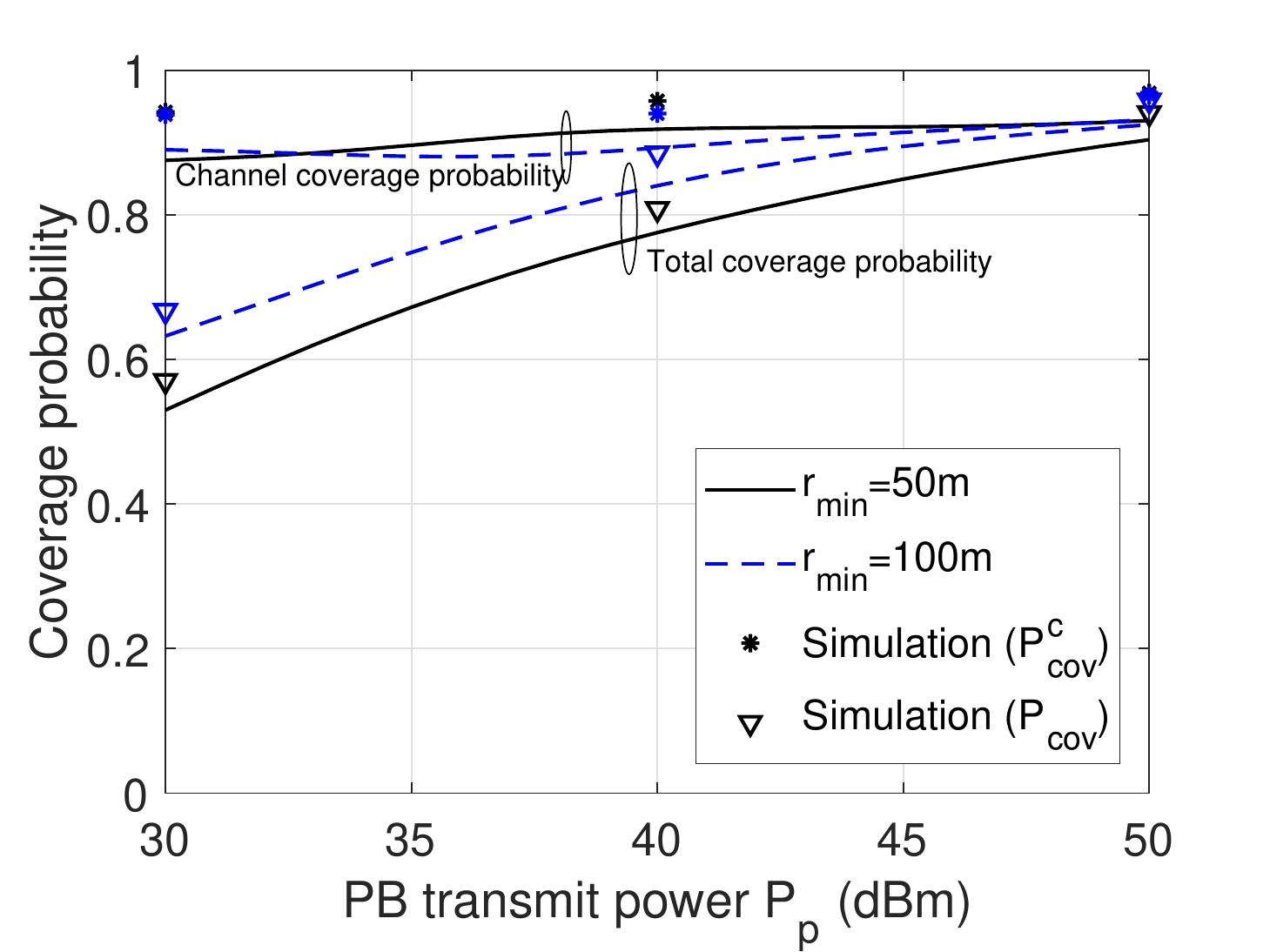}\label{fig:Pp}}
\subfigure[Total coverage and asymptotic total coverage.]{\includegraphics[scale=0.6]{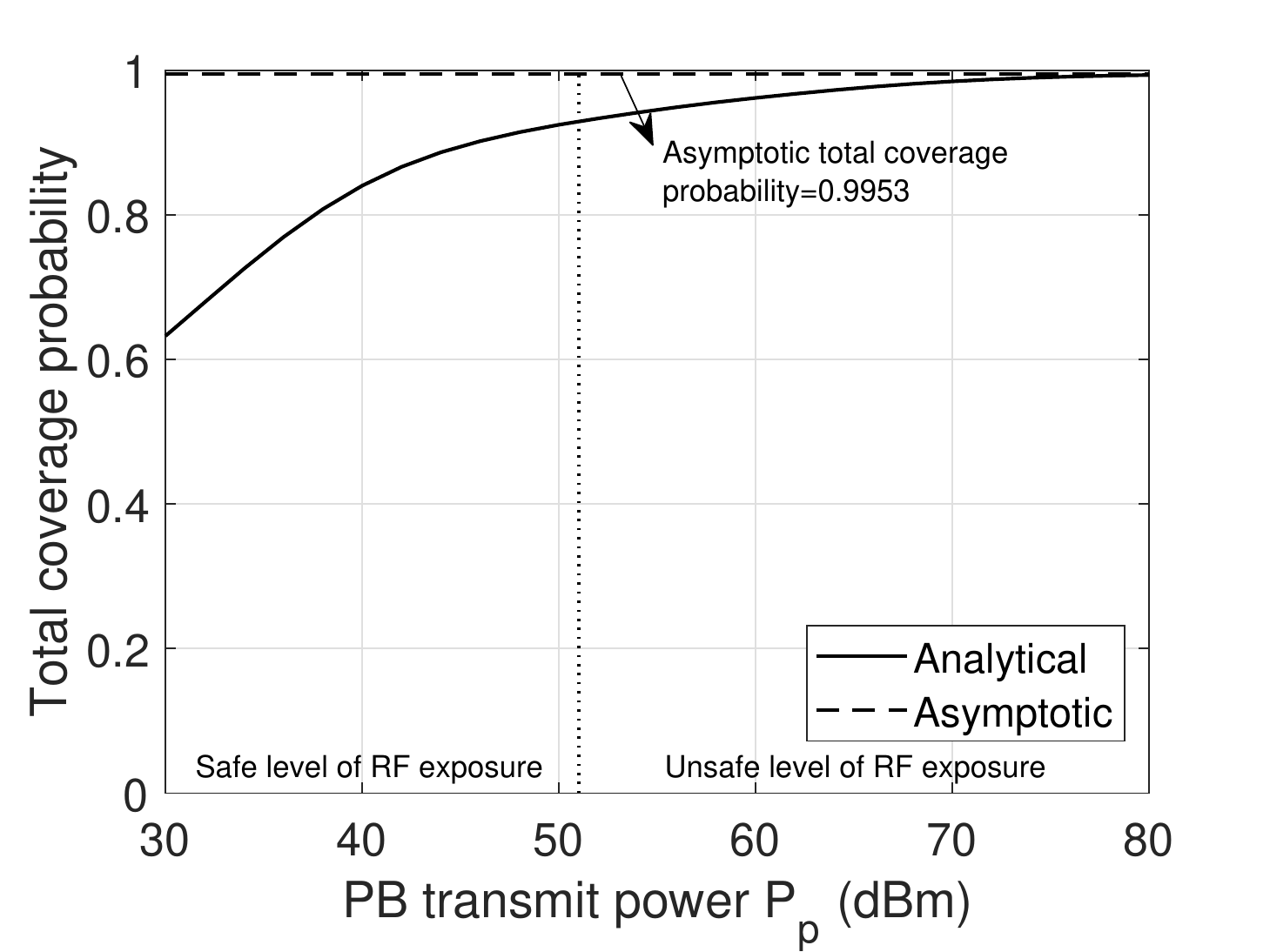}\label{fig:asym}}
\vspace{-3mm}
\caption{Coverage probabilities versus PB transmit power $P_p$.}
\end{figure*}

Fig.~\ref{fig:Pp} illustrates the effect of PB transmit power $P_p$ on the total coverage probability and channel coverage probability, with different radius of the LOS region $\rmin=50 \textrm{m}, 100 \textrm{m}$. The simulation results are also plotted in the figure, which are averaged over $10^8$ Monte Carlo simulation runs. The accuracy is between $3\%$ to $8\%$, which again validates the proposed model. Hence, in the subsequent figures in the paper we only show the analytical results and discuss the insights.

\redcom{Fig.~\ref{fig:asym} plots the total coverage probability against the transmit power of PB. We also plot an asymptotic result when $P_p$ approaches infinity. This result is obtained as follows. As $P_p$ approaches infinity, if one or more PBs fall into the LOS or NLOS region of a TX, this TX will be active and transmit with a power of $P_t=\min(\frac{\rho}{1-\rho}\Pmax_1,\Pmax_2)$. Hence, the asymptotic power coverage probability is equivalent to the probability that at least one PB falls into the LOS or NLOS region of the TX, which is given by
\begin{align}
\lim_{P_p\to\infty}\Pc^P=1-\!\exp\!\left(-\pi\lambda_p\rmax^2\right).
\end{align}
The asymptotic conditional channel coverage probability and the asymptotic total coverage probability can be found by \eqref{eq:pcout_rlpg} and \eqref{eq:pout} respectively with the portion of TXs at the $n$th level as
\begin{align}
\lim_{P_p\to\infty}k_n=\begin{cases}
				0, & n=\{0,1,2,...,N-1\}\\
				\Pc^P, & n=N
		\end{cases}.
\end{align}
From the figure, we can see that the analytical and asymptotic results converge as $P_p$ gets large, which validates the derivation of the asymptotic results. In addition, in Fig.~\ref{fig:asym}, we have marked the safe RF exposure region with a PB transmit power less than 51 dBm, equivalently power density smaller than 10 $\textrm{W/m}^2$ at 1 m from the PB~\cite{Xia-2015}. We will discuss in detail later in the feasibility study in Section~\ref{sec:feasible}.}

\textit{Insights:} Fig.~\ref{fig:Pp} shows that: (i) The channel coverage probability first slightly decreases and then increases with the increase of $P_p$. This can be explained as follows. \redcom{At first, both the transmit power of the desired TX and the number of interfering TX increase with $P_p$. The interplay of this two factors results in the slightly decreasing trend for the channel coverage probability. As $P_p$ further increases, the increase in the number of interfering TX is negligible, while the transmit power of the desired TX continues to increase, which leads to the increase of the channel coverage probability.} (ii) The total coverage probability increases as PB transmit power $P_p$ increases. When $P_p$ is small, the desired TX might not receive enough power to activate the IT process. So the total coverage probability is small and is limited by the power coverage probability. When $P_p$ is large, the channel coverage probability becomes the dominant factor in determining the total coverage probability. Hence, eventually the channel coverage probability and total coverage probability curves merge. (iii) The total coverage probability increase with $\rmin$, because more PBs falls into the LOS region and the path-loss is less severe, which improves the power coverage probability. The benefit of increasing the radius of the LOS region is less significant for the channel coverage probability.



\subsection{Effect of Directional Beamforming at PB, TX and RX}\label{sec:gammaPT}
\begin{figure*}[t]
\begin{minipage}[t]{0.48\linewidth}
\centering
\includegraphics[width=1 \textwidth]{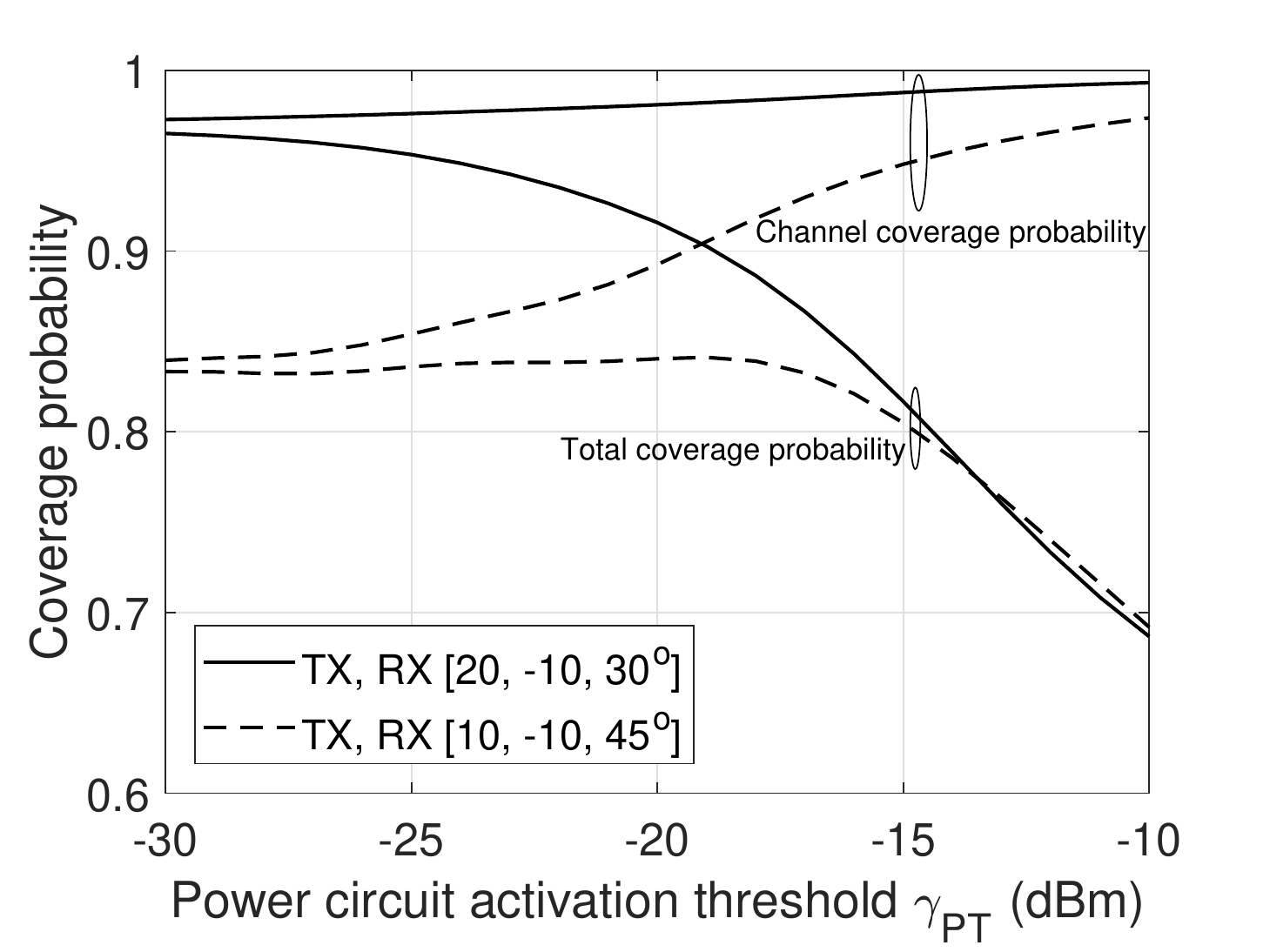}
\centering
\vspace{-12mm}
\caption{Channel coverage probability and total coverage probability versus power circuit activation threshold $\gammapt$ with different TX and RX beamforming parameters.}
\centering
\label{fig:gammaPT}
\end{minipage}\hfill
\centering
\begin{minipage}[t]{0.48\linewidth}
\centering
\includegraphics[width=1 \textwidth]{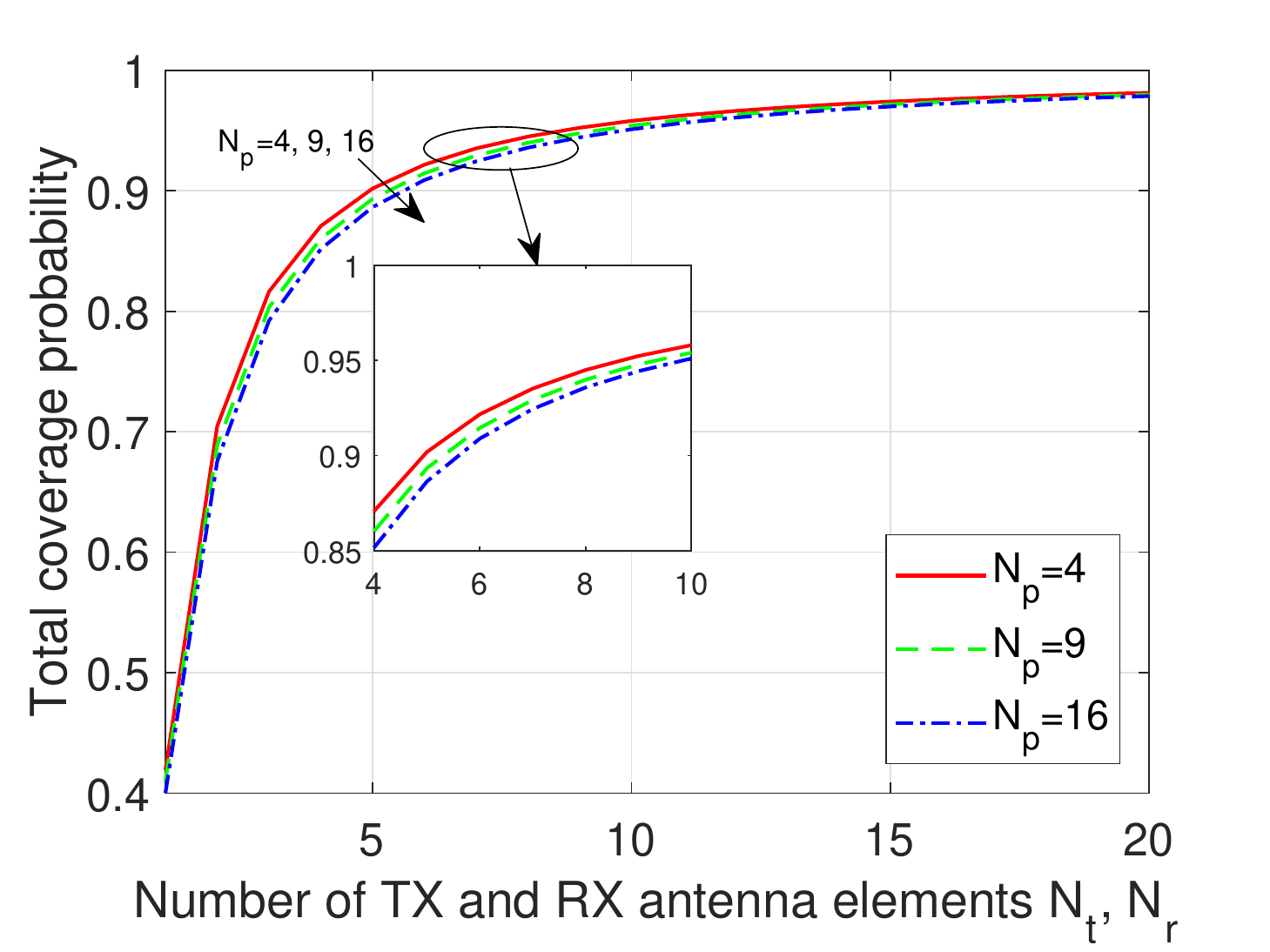}
\centering
\vspace{-12mm}
\caption{Total coverage probability versus the numbers of antenna elements at TX and RX $\mathcal{N}_t$ and $\mathcal{N}_r$ with different numbers of antenna elements at PB.}
\centering
\label{fig:Np}
\end{minipage}
\end{figure*}

Fig.~\ref{fig:gammaPT} plots the total coverage probability and channel coverage probability against the power circuit activation threshold of TX for different beamforming parameters at TX and RX, i.e., [20 dB, $-$10 dB, $30^\textrm{o}$] and [10 dB, $-$10 dB, $45^\textrm{o}$].

\textit{Insights:} Fig.~\ref{fig:gammaPT} shows that, for both sets of beamforming parameters, as the power circuit activation threshold $\gammapt$ increases, the channel coverage probability is always increasing, while the total coverage probability stays roughly the same at first and then decreases. This can be explained as follows. When $\gammapt$ increases, the power coverage probability decreases. The reduction in the number of active TXs improves the channel coverage probability. With regards to the total coverage probability, its trend is determined by the interplay of channel coverage probability and power coverage probability. At first, the drop in power coverage is relatively small as shown in Fig.~\ref{fig:ppout}; so the total coverage probability is almost unchanged. After a certain point, the power coverage probability drops a lot, which mainly governs the total coverage probability. Hence, the total coverage probability decreases later on.

Comparing the curves for the different beamforming parameters, we can see that TX and RX with [20 dB, $-$10 dB, $30^\textrm{o}$] gives a higher total coverage probability in the low power circuit activation threshold region. This is because a narrower main lobe beam-width gives a larger main lobe gain and makes less interfering TXs fall into its main lobe which results in higher channel coverage probability. However, the total coverage probability is limited by the power coverage probability when $\gammapt$ is large.

\textit{Impact of number of antenna elements:} The beamforming model adopted in this paper can be related to any specific array geometry by substituting the appropriate values for the three beamforming parameters. For instance, a uniform planar square array with half-wavelength antenna element spacing can be used at the PBs, TXs and RXs. The values for the main lobe antenna gain $\Gmax_a$, side lobe antenna gain $\Gmin_a$ and main lobe beamwidth $\theta_a$ depend on the number of the antenna elements $\mathcal{N}_a$ and can be calculated by using the equations below~\cite{Venugopal-2015c}:
\ifCLASSOPTIONpeerreview
\begin{align}
\Gmax_a=\mathcal{N}_a,\Gmin_a=\frac{\sqrt{\mathcal{N}_a}-\frac{\sqrt{3}}{2\pi}\mathcal{N}_a\sin(\frac{\sqrt{3}}{2\sqrt{\mathcal{N}_a}})}{\sqrt{\mathcal{N}_a}-\frac{\sqrt{3}}{2\pi}\sin(\frac{\sqrt{3}}{2\sqrt{\mathcal{N}_a}})},
\theta_a=\frac{\sqrt{3}}{\sqrt{\mathcal{N}_a}},
\end{align}
\else
\begin{align}
\Gmax_a=\mathcal{N}_a,
\end{align}
\begin{align}
\Gmin_a=\frac{\sqrt{\mathcal{N}_a}-\frac{\sqrt{3}}{2\pi}\mathcal{N}_a\sin(\frac{\sqrt{3}}{2\sqrt{\mathcal{N}_a}})}{\sqrt{\mathcal{N}_a}-\frac{\sqrt{3}}{2\pi}\sin(\frac{\sqrt{3}}{2\sqrt{\mathcal{N}_a}})},
\end{align}
\begin{align}
\theta_a=\frac{\sqrt{3}}{\sqrt{\mathcal{N}_a}},
\end{align}
\fi

\noindent where subscript $a=p$ for PB, $a=t$ for TX and $a=r$ for RX.

Fig.~\ref{fig:Np} plots the total coverage probability versus the numbers of antenna elements at the TX and RX $\mathcal{N}_t$ and $\mathcal{N}_r$ with different PB antenna element number $\mathcal{N}_p$. The figure shows that the total coverage probability increases with the numbers of antenna elements at the TX and RX, which agrees with our previous findings. However, \redcom{under our considered system parameters, the total coverage probability stays roughly the same after having more than about 15 TX and RX antenna elements, as the side lobe antenna gain and the main lobe beamwidth stay almost constant with further increase in the number of antenna elements.} In addition, the number of antenna elements at the PB does not significantly impact the total coverage probability.
%
%


\subsection{Effect of Allowed Maximum Harvested Power at TX}\label{sec:Pmax1}
\begin{figure*}[t]
\begin{minipage}[t]{0.48\linewidth}
\centering
\includegraphics[width=1 \textwidth]{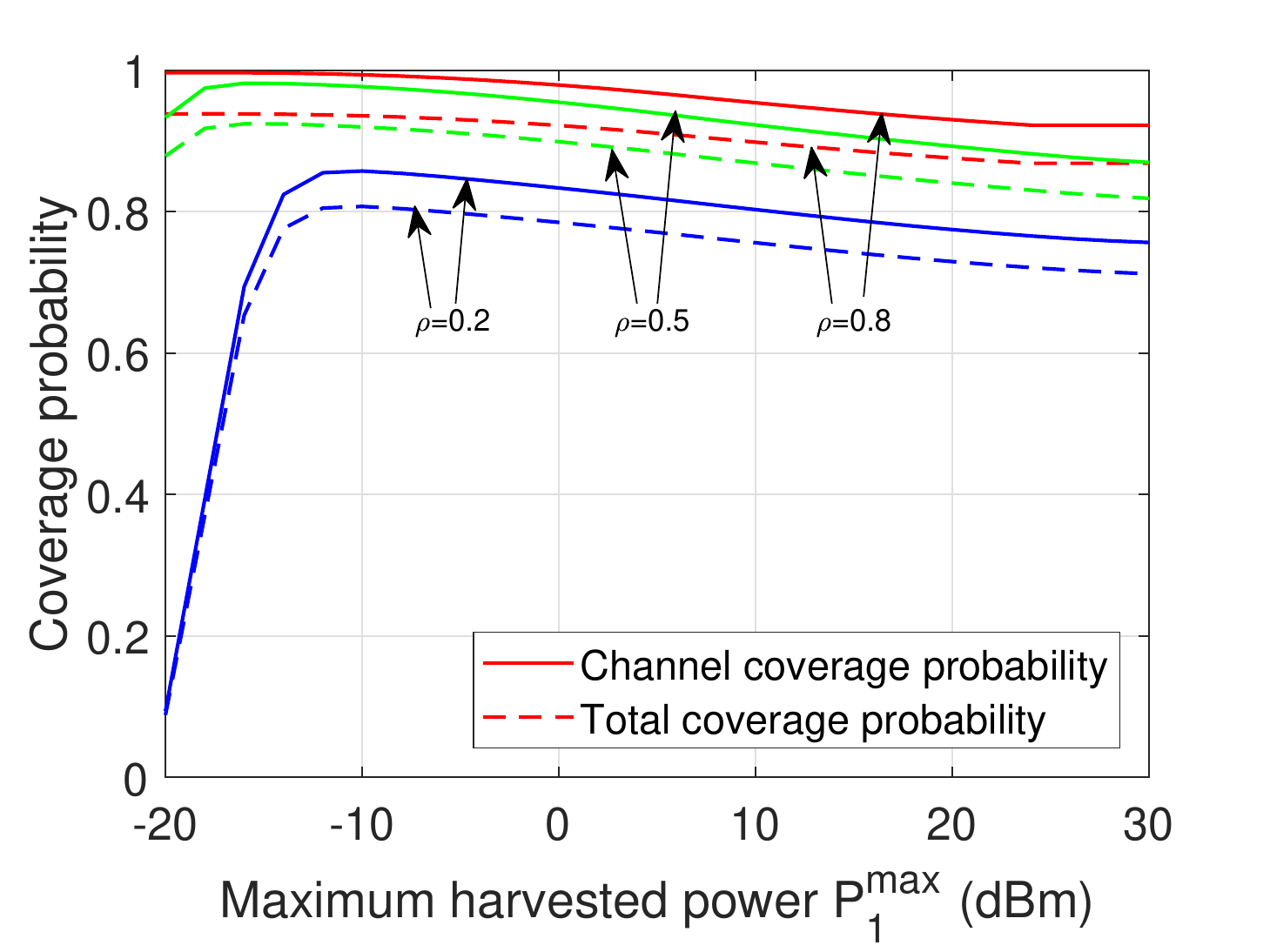}
\centering
\vspace{-12mm}
\caption{Channel coverage probability and total coverage probability versus allowed maximum harvested power $\Pmax_1$ with different time switching ratios.}
\centering
\label{fig:Pmax1}
\end{minipage}\hfill
\centering
\begin{minipage}[t]{0.48\linewidth}
\centering
\includegraphics[width=1 \textwidth]{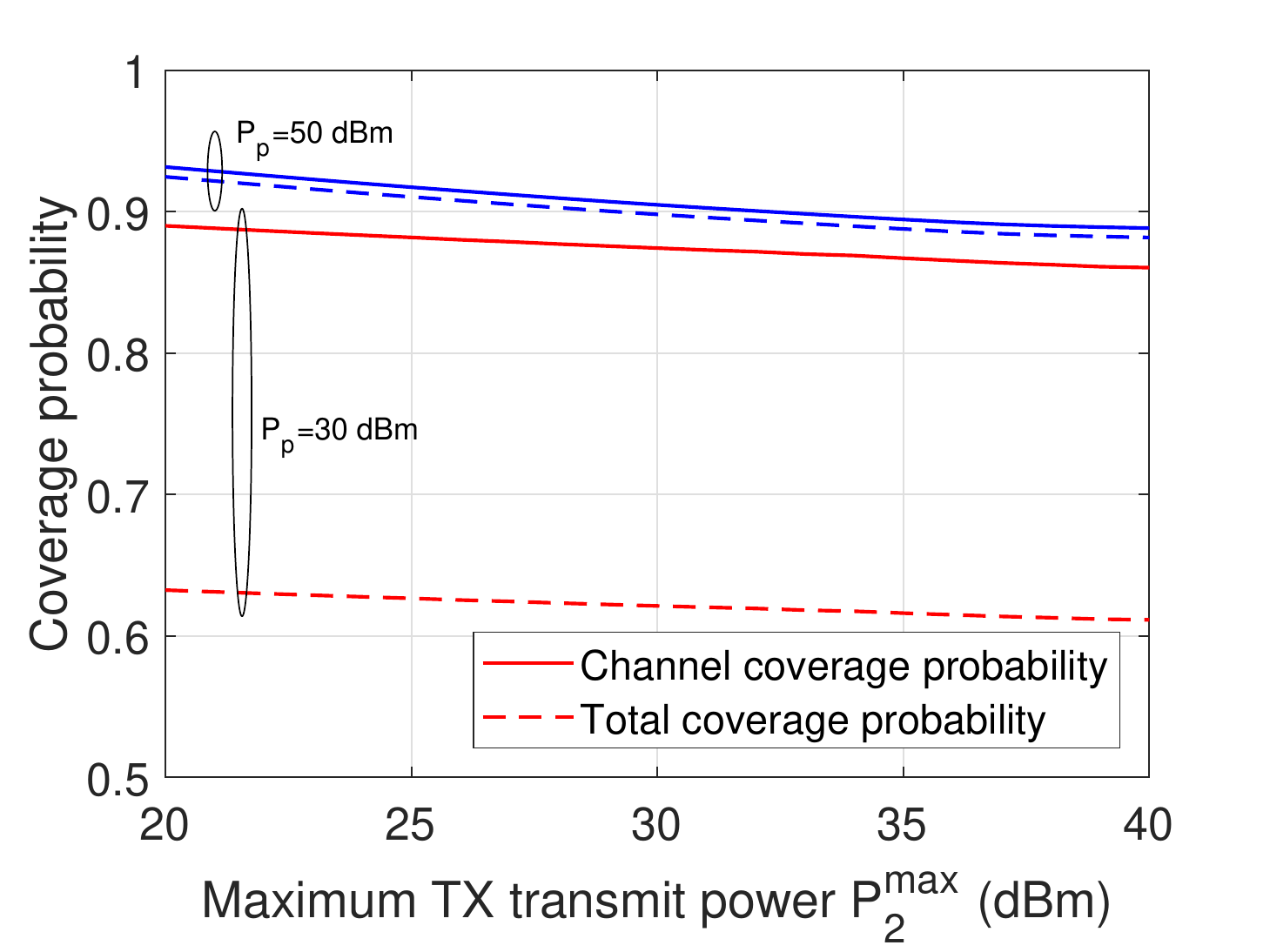}
\centering
\vspace{-12mm}
\caption{Channel coverage probability and total coverage probability versus maximum TX transmit power $\Pmax_2$ for different PB transmit power with the allowed maximum harvested power of TX being 50 dBm.}
\centering
\label{fig:Pmax2}
\end{minipage}
\end{figure*}

Fig.~\ref{fig:Pmax1} plots the total coverage probability and channel coverage probability against the allowed maximum harvested power of TX $\Pmax_1$ for different time switching ratios $0.2$, $0.5$ and $0.8$. Note that both the time switching ratio and the allowed maximum harvested power do not affect the power coverage probability.

\textit{Insights:} Fig.~\ref{fig:Pmax1} shows that the channel coverage probability and the total coverage probability both  first increase with $\Pmax_1$, then decrease. The rise of the channel coverage probability is because the possible transmit power of the desired TX increases with its allowed maximum harvested power. However, as $\Pmax_1$ further increases, the accumulated harvested energy during the PT phase is higher and the transmit power of other active TX also goes up. As a result, the interfering power received at the RX is higher and the channel coverage probability decreases. The channel coverage probability will converge to a constant value as $\Pmax_1$ increases even further, because the maximum transmit power of active TX has limited the channel performance.

Comparing the curves for different $\rho$, we can see that for a given maximum harvested power of TX $\Pmax_1$, increasing $\rho$ improves the coverage probabilities. When $\rho$ is higher, more energy is captured during the PT phase. Therefore, the transmit power of active TX is now limited by the maximum transmit power $\Pmax_2$. As a result, the channel coverage probability and total coverage probability converge and do not vary much with the changes in the allowed maximum harvested power.

\subsection{Feasibility of PB-assisted mmWave Wireless Ad hoc Networks}\label{sec:feasible}
Finally, we investigate the feasibility of PB-assisted mmWave wireless ad hoc network. Fig.~\ref{fig:Pmax2} is a plot of the total coverage probability and channel coverage probability versus maximum TX transmit power $\Pmax_2$ with varied PB transmit power, $50$ dBm and $30$ dBm. To better highlight the impact of the maximum transmit power at TX, we have set $\Pmax_1$ equal to $50$ dBm which is much higher than $\Pmax_2$. From the figure, we can see that the channel coverage probability and total coverage probability do not change much with the considered maximum TX transmit power, which means that the probability mass function (PMF) of the transmit power for the desired TX remains almost the same. Note that the power coverage probability is independent of the maximum transmit power of TX.

\textit{Insight:} From Fig.~\ref{fig:Pmax2}, the total coverage probability and the channel coverage probability are around $90\%$ if $P_p$ is $50$ dBm. If PB transmits with a constant power of $50$ dBm, the power density at a distant of $1$ m from the PB is $7.95$ $\textrm{W}/\textrm{m}^2$. This power density is smaller than $10$ $\textrm{W}/\textrm{m}^2$, which is the permissible safety level of human exposure to RF electromagnetic fields based on IEEE Standard. \redcom{Under this safety regulation, the maximum permissible PB transmit power would be 51 dBm. We have marked this value in Fig.~\ref{fig:asym}. From Fig.~\ref{fig:asym}, we can see that the total coverage probability with a PB transmit power less than 51 dBm can be up to 93.4\% of the maximum system performance, as given by the asymptotic analysis in Section~\ref{sec:Pp}, based on our considered system parameters. The results in Fig.~\ref{fig:asym} and Fig.~\ref{fig:Pmax2} show that PB-assisted mmWave ad hoc networks are feasible under practical network setup.}

\section{Conclusion}\label{sec:conclusion}
In this paper, we have presented an approximate yet accurate model for PB-assisted mmWave wireless ad hoc networks, where TXs harvest energy from all PBs and then use the harvested energy to transmit information to their corresponding RXs. We first obtained the Laplace transform of the aggregate received power at the TX to compute the power coverage probability. Then, the channel coverage probability and total coverage probability were formulated based on discretizing the transmit power of TXs into a finite number of levels. The simulation results confirmed the accuracy of the proposed model. The results have shown that the total coverage probability improves by increasing the transmit power of PB, narrowing the main lobe beam-width and decreasing the maximum harvested power at the TX. Our results also showed that PB-assisted mmWave ad hoc network is feasible under realistic setup conditions. \redcom{Future work can consider extensions to other MAC protocols such as carrier-sense multiple access (CSMA)~\cite{Elsway-2013,Kaynia-2011} and optimal allocation of the transmit power of an active TX.}

\appendices
\section{Proof of Theorem \ref{th:laplace_Ppt}}\label{app:1}
Following the definition of Laplace transform, the Laplace transform of the aggregate received power can be expressed as
\begin{align}
&\mathcal{L}_{\Ppt}(s)=\mathbb{E}_{\Ppt}\left[\exp(-s\Ppt)\right]=\mathbb{E}_{\phi_p,G_{i0},g_{i0}}\left[\exp\left(-sP_p\sum_{Z_i\in\phi_p}{G_{i0}g_{i0}l(r_i)}\right)\right]\nonumber\\
&=\mathbb{E}_{\phi_p,G_{i0},g_{i0}}\left[\exp\left(-sP_p\sum_{0\leqslant r_i<1}{G_{i0}g_{i0}l(r_i)}\right)\right]\mathbb{E}_{\phi_p,G_{i0},g_{i0}}\left[\exp\left(-sP_p\sum_{1\leqslant r_i<\rmin}{G_{i0}g_{i0}l(r_i)}\right)\right]\nonumber\\
&\times\mathbb{E}_{\phi_p,G_{i0},g_{i0}}\left[\exp\left(-sP_p\sum_{\rmin\leqslant r_i<\rmax}{G_{i0}g_{i0}l(r_i)}\right)\right]\nonumber\\ 
&=\underbrace{\exp\!\!\left(-\!\int_{-\pi}^{\pi}\int_0^1{\mathbb{E}_{G_{i0},g_{i0}}[1-\exp(-sP_pG_{i0}g_{i0})]\lambda_p rdrd\theta}\right)}_{A_1}\nonumber\\
&\times\underbrace{\exp\!\!\left(\!\!-\!\!\int_{-\!\pi}^{\pi}\!\int_1^{\rmin}{\!\!\mathbb{E}_{G_{i0},g_{i0}}[1\!-\!\exp(\!-\!sP_pG_{i0}g_{i0}r^{-\!\alphaL}\!)]\lambda_p rdrd\theta}\!\right)}_{A_2}\nonumber\\
&\times\underbrace{\exp\!\!\left(\!\!-\!\!\int_{-\!\pi}^{\pi}\!\int_{\rmin}^{\rmax}{\!\!\mathbb{E}_{G_{i0},g_{i0}}[1\!-\!\exp(\!-\!sP_pG_{i0}g_{i0}\beta r^{-\!\alphaN}\!)]\lambda_p rdrd\theta}\!\right)}_{A_3}\label{eq:proofmgf}.
\end{align}

The first term $A_1$ is evaluated as follows
\ifCLASSOPTIONpeerreview
\begin{align}
A_1&=\exp\left(-\pi\lambda_p\left(1-\mathbb{E}_{G_{i0},g_{i0}}[\exp(-sP_pG_{i0}g_{i0})]\right)\right)\nonumber\\
&=\exp\left(-\pi\lambda_p\left(1-\mathbb{E}_{G_{i0}}\left[\int_0^\infty\exp(-sP_pG_{i0}g)f_{\gL}(g)dg\right]\right)\right)\nonumber\\
&=\exp\left(-\pi\lambda_p+\pi\lambda_p m^m\mathbb{E}_{G_{i0}}\left[(m+sP_pG_{i0})^{-m}\right]\right)\nonumber\\
&=\exp\left(-\pi\lambda_p+\pi\lambda_p m^m\sum_{k=1}^4(m+sP_pG_k)^{-m}p_k\right),
\end{align}
\else
\begin{align}
A_1&=\exp\left(-\pi\lambda_p\left(1-\mathbb{E}_{G_{i0},g_{i0}}[\exp(-sP_pG_{i0}g_{i0})]\right)\right)\nonumber\\
&=\exp\left(-\pi\lambda_p\left(1-\mathbb{E}_{G_{i0}}\left[\int_0^\infty\exp(-sP_pG_{i0}g)f_{\gL}(g)dg\right]\right)\right)\nonumber\\
&=\exp\left(-\pi\lambda_p+\pi\lambda_p m^m\mathbb{E}_{G_{i0}}\left[(m+sP_pG_{i0})^{-m}\right]\right)\nonumber\\
&=\exp\left(-\pi\lambda_p+\pi\lambda_p m^m\sum_{k=1}^4(m+sP_pG_k)^{-m}p_k\right),
\end{align}
\fi

\noindent where we use the fact that the link in LOS state experiences Nakagami-$m$ fading with $f_{\gL}(g)=\frac{m^mg^{m-1}\exp(-mg)}{\Gamma(m)}$.

The second term $A_2$ is evaluated as follows
\begin{subequations}
\begin{align}
A_2&=\exp\Big(\pi\lambda_p\mathbb{E}_{G_{i0},g_{i0}}\left[1-\exp(-sP_pG_{i0}g_{i0})\right]-\pi\lambda_p\rmin^2\mathbb{E}_{G_{i0},g_{i0}}\left[1-\exp(-s\rmin^{-\alphaL}P_pG_{i0}g_{i0})\right]\nonumber\\
&-\pi\lambda_p\mathbb{E}_{G_{i0},g_{i0}}\left[(sP_pG_{i0})^{\deltaL}g_{i0}^{\deltaL}\gamma(1-\deltaL,sP_pg_{i0}G_{i0})\right]\nonumber\\
&\left.+\pi\lambda_p\mathbb{E}_{G_{i0},g_{i0}}\left[(sP_pG_{i0})^{\deltaL}g_{i0}^{\deltaL}\gamma(1-\deltaL,sP_pg_{i0}G_{i0}\rmin^{-\alphaL})\right]\right)\label{eq:proofa2_1}
\end{align}
\begin{align}
&=\exp\left(\pi\lambda_p\!-\!\pi\lambda_p m^m\sum_{k=1}^4(m+sP_pG_k)^{-m}p_k\!-\!\pi\lambda_p\rmin^2+\sum_{k=1}^4\pi\lambda_p\rmin^2 m^m(m+s\rmin^{-\alphaL}P_pG_k)^{-m}p_k\right.\nonumber\\
&+\pi\lambda_p\sum_{k=1}^4\!\left(sP_pG_k\right)^{\deltaL}\frac{m^m(sP_pG_k)^{-\!\deltaL\!-\!m}\alphaL\Gamma(1\!+\!m)}{(2+m\alphaL)\Gamma(m)}\,_2F_1\left(\!1+m,\!m+\deltaL;\!1+m+\deltaL;\!-\frac{m}{sP_pG_k}\!\right)p_k\nonumber\\
&-\pi\lambda_p\sum_{k=1}^4\!\left(sP_pG_k\right)^{\deltaL}\frac{m^m(\rmin^{-\alphaL}sP_pG_k)^{-\!\deltaL\!-\!m}\alphaL\Gamma(1\!+\!m)}{(2+m\alphaL)\Gamma(m)}\nonumber\\
&\left.\times\,_2F_1\left(\!1+m,\!m+\deltaL;\!1+m+\deltaL;\!-\frac{\rmin^{\alphaL}m}{sP_pG_k}\right)p_k\right)\label{eq:proofa2_2},
\end{align}
\end{subequations}

\noindent where \eqref{eq:proofa2_1} follows from changing variables and integration by parts and \eqref{eq:proofa2_2} is obtained after taking the expectation over $\gL$ then $G_{i0}$.

Similarly, the third term $A_3$ can be worked out by taking the expectation over $\gN$, which has a PDF as $f_{\gN}(h)=\exp(-g)$. The details are omitted for sake of brevity. Finally, the Laplace transform in Theorem~\ref{th:laplace_Ppt} is obtained by substituting $A_1$, $A_2$ and $A_3$ into \eqref{eq:proofmgf}.

\section{Proof of Proposition \ref{pr:pcout}}\label{app:2}
By substituting \eqref{eq:sinr} into \eqref{eq:pcout}, we can express the conditional channel coverage probability as
\ifCLASSOPTIONpeerreview
\begin{subequations}
\begin{align}
\Pc^C(\gammatr)=&\Pr\left(\frac{P_{X_0} D_0 h_0 l(d_0)}{\sum_{X_i\in\phi_\textrm{active}}P_{X_i} D_{i0} h_{i0} l(X_i)+\sigma^2}>\gammatr\right)\nonumber\\
\approx&\Pr\left(\frac{P_{X_0} D_0 h_0 l(d_0)}{\sum_{n=0}^N\sum_{X_i\in\phi_t^n}P_t^n D_{i0} h_{i0} l(X_i)+\sigma^2}>\gammatr\right)\label{eq:pr2a}\\
=&\Pr\left(h_0>\frac{\gammatr(I_X+\sigma^2)}{P_{X_0} D_0 l(d_0)}\right)=\mathbb{E}_{P_{X_0}, I_X}\left[1-F_{h_0}\left(\frac{\gammatr(I_X+\sigma^2)}{P_{X_0} D_0 l(d_0)}\right)\right],\label{eq:pr2b}
\end{align}
\end{subequations}
\else
\begin{subequations}
\begin{align}
\Pc^C(\gammatr)=&\Pr\left(\frac{P_{X_0} D_0 h_0 l(d_0)}{\sum_{X_i\in\phi_\textrm{active}}P_{X_i} D_{i0} h_{i0} l(X_i)+\sigma^2}>\gammatr\right)\nonumber\\
\approx&\Pr\left(\frac{P_{X_0} D_0 h_0 l(d_0)}{\sum_{n=0}^N\sum_{X_i\in\phi_t^n}P_t^n D_{i0} h_{i0} l(X_i)+\sigma^2}>\gammatr\right)\label{eq:pr2a}\\
=&\Pr\left(h_0>\frac{\gammatr(I_X+\sigma^2)}{P_{X_0} D_0 l(d_0)}\right)\nonumber\\
=&\mathbb{E}_{P_{X_0}, I_X}\left[1-F_{h_0}\left(\frac{\gammatr(I_X+\sigma^2)}{P_{X_0} D_0 l(d_0)}\right)\right],\label{eq:pr2b}
\end{align}
\end{subequations}
\fi

\noindent where approximation in \eqref{eq:pr2a} comes from our power level discretization, $I_X=\sum_{n=0}^N\sum_{X_i\in\phi_t^n}P_t^n D_{i0} h_{i0} l(X_i)$ and $F_{h_0}(\cdot)$ is the CDF of the fading power gain on the reference TX-RX link. Since the desired link is assumed to experience Nakagami-$m$ fading with integer $m$, the CDF of $h_0$ has a nice form, which is $F_{h_0}(h)=1-\sum_{l=0}^{m-1}\frac{1}{l!}(mh)^l\exp(-mh)$. Hence, we can re-write \eqref{eq:pr2b} as
\begin{align}
\Pc^C(\gamma_{\mathrm{TR}})=&\mathbb{E}_{P_{X0}, I_X}\left[\sum_{l=0}^{m-1}\frac{1}{l!}\left(m\frac{\gammatr(I_X+\sigma^2)}{P_{X_0} D_0 l(d_0)}\right)^l\exp\left(-m\frac{\gammatr(I_X+\sigma^2)}{P_{X_0} D_0 l(d_0)}\right)\right]\nonumber\\
=&\sum_{n=0}^N\sum_{l=0}^{m-1}\frac{1}{l!}\mathbb{E}_{I_X}\left[\left(m\frac{\gammatr(I_X+\sigma^2)}{P_t^n D_0 l(d_0)}\right)^l\exp\left(-m\frac{\gammatr(I_X+\sigma^2)}{P_t^n D_0 l(d_0)}\right)\right]\frac{k_n}{\Pc^P(\gammapt)}\label{eq:pr2c},
\end{align}
where the PMF of $P_{X_0}$ is $\Pr(P_{X_0}=P_t^n)=\frac{k_n}{\Pc^P(\gammapt)}$ in \eqref{eq:pr2c}, as we assume that the desired TX is active.

The general form of the Laplace transform of $I_X+\sigma^2$ is $\mathcal{L}_{I_X+\sigma^2}(s)=\mathbb{E}_{I_X}[\exp(-s(I_X+\sigma^2))]$. Taking $l$th derivative with respect to $s$, we achieve
\begin{align}\label{eq:pr2d}
\frac{dl}{ds^l}\mathcal{L}_{I_X+\sigma^2}(s)=\mathbb{E}_{I_X}\left[\frac{dl}{ds^l}\exp(-s(I_X+\sigma^2))\right]=\mathbb{E}_{I_X}\left[(-I_X-\sigma^2)^l\exp(-s(I_X+\sigma^2))\right].
\end{align}

Comparing \eqref{eq:pr2d} with the expectation term in \eqref{eq:pr2c}, we have
\begin{align}
\Pc^C(\gamma_{\mathrm{TR}})=&\sum_{n=0}^N\sum_{l=0}^{m-1}\frac{(-s)^l}{l!}\frac{dl}{ds^l}\mathcal{L}_{I_X+\sigma^2}(s)\frac{k_n}{\Pc^P(\gammapt)},
\end{align}
where $s=\frac{m\gammatr}{P_t^nD_0l(d_0)}$. Hence, we arrive the result in Proposition~\ref{pr:pcout}.

\bibliographystyle{IEEEtran}

\end{document}